\documentclass[aps,prx, twocolumn,showpacs,preprintnumbers,superscriptaddress]{revtex4-2}
\usepackage{graphicx}  % Include figure files
\usepackage{subfigure}
\usepackage{multirow}

\usepackage{fancyhdr}
\usepackage{longtable}
\usepackage{parskip}
\usepackage[T1]{fontenc}
\usepackage{dcolumn}   % Align table columns on decimal point
\usepackage{svg}
\usepackage{bm}        % bold math
\usepackage{amsfonts}  % Common math fonts
\usepackage{amsmath}   % Common math functions
\usepackage{amssymb}   % Common math symbols
\newcommand{\pwisein}{\left\{ \begin{array}{ll}}
\newcommand{\pwiseout}{\end{array}\right.}

\begin{document}

\title{Optimal filtering and generation of entangled photons for quantum applications in the presence of noise
}

\newcommand{\equalcontrib}{These authors contributed equally to this work.}

\author{Jordan M. Thomas$^{\dagger}$}   % dagger added manually
\thanks{jmthomas@fnal.gov. Current address: Fermi National Accelerator Laboratory, Batavia, IL 60510, USA}
\affiliation{Center for Photonic Communication and Computing, ECE Department, Northwestern University, 2145 Sheridan Road, Evanston, Illinois 60208, USA}

\author{Andrew R. Cameron}\thanks{\equalcontrib}
\affiliation{Fermi National Accelerator Laboratory, Batavia, IL 60510, USA}

\author{Akil Pathiranage}
\affiliation{Fermi National Accelerator Laboratory, Batavia, IL 60510, USA}
\author{Si Xie}
\affiliation{Fermi National Accelerator Laboratory, Batavia, IL 60510, USA}
\affiliation{Division of Physics, Mathematics and Astronomy, California Institute of Technology, Pasadena, CA 91125, USA}
\author{Raju Valivarthi}
\affiliation{Division of Physics, Mathematics and Astronomy, California Institute of Technology, Pasadena, CA 91125, USA}
\affiliation{Alliance for Quantum Technologies (AQT), California Institute of Technology, Pasadena, CA 91125, USA}
\author{Panagiotis Spentzouris}
\affiliation{Fermi National Accelerator Laboratory, Batavia, IL 60510, USA}
\author{Maria Spiropulu}
\affiliation{Division of Physics, Mathematics and Astronomy, California Institute of Technology, Pasadena, CA 91125, USA}
\affiliation{Alliance for Quantum Technologies (AQT), California Institute of Technology, Pasadena, CA 91125, USA}
\author{Cristi\'{a}n Pe\~{n}a}
\affiliation{Fermi National Accelerator Laboratory, Batavia, IL 60510, USA}
\author{Prem Kumar}
\affiliation{Center for Photonic Communication and Computing, ECE Department, Northwestern University, 2145 Sheridan Road, Evanston, Illinois 60208, USA}\affiliation{Department of Physics and Astronomy, Northwestern University, 2145 Sheridan Road, Evanston, Illinois 60208, USA}

%\author{Andrew R. Cameron}
%\affiliation {\it Fermilab}

%\date{\today}

\begin{abstract} 
Filtering is commonly used in quantum optics to reject noise photons, and also to enable interference between independent photons. However, filtering the joint spectrum of photon pairs can reduce the inherent coincidence probability or loss-independent heralding efficiency. Here, we investigate filtering for multiphoton applications based on entanglement and interference (e.g., quantum teleportation). We multiplex C-band entangled photons and C-band classical communications into the same long-distance fibers, which enables scalable low-loss quantum networking but requires filtering of spontaneous Raman scattering noise from classical light. Using tunable-bandwidth filters, low-jitter detectors, and polarization filters, we co-propagate time-bin-entangled photons at wavelengths compatible with erbium-ion quantum memories (1536.5\,nm) and 10-Gbps C-band classical data over 25\,km/25\,km of standard fiber. Narrow filtering enables mW-level C-band power, which exceeds comparable studies by roughly an order of magnitude and could feasibly support Tbps classical rates. We evaluate how performance depends on pump and filter bandwidths, multipair emission, filter shapes, loss, phase matching, and how quantum information is measured. We find a trade-off between improving noise impact and single-mode purity and discuss mitigation methods toward optimal multiphoton applications. Importantly, these results apply to noise in free space and in quantum devices (sources, frequency converters, switches, detectors, etc.) and provide insight on filter-induced degradation of single-photon purity and rates even in noise-free environments.
\end{abstract}
\maketitle

\section{Introduction}
\begin{figure*}[http]
\centering
%\onecolumn
\includegraphics[width=1\linewidth]
{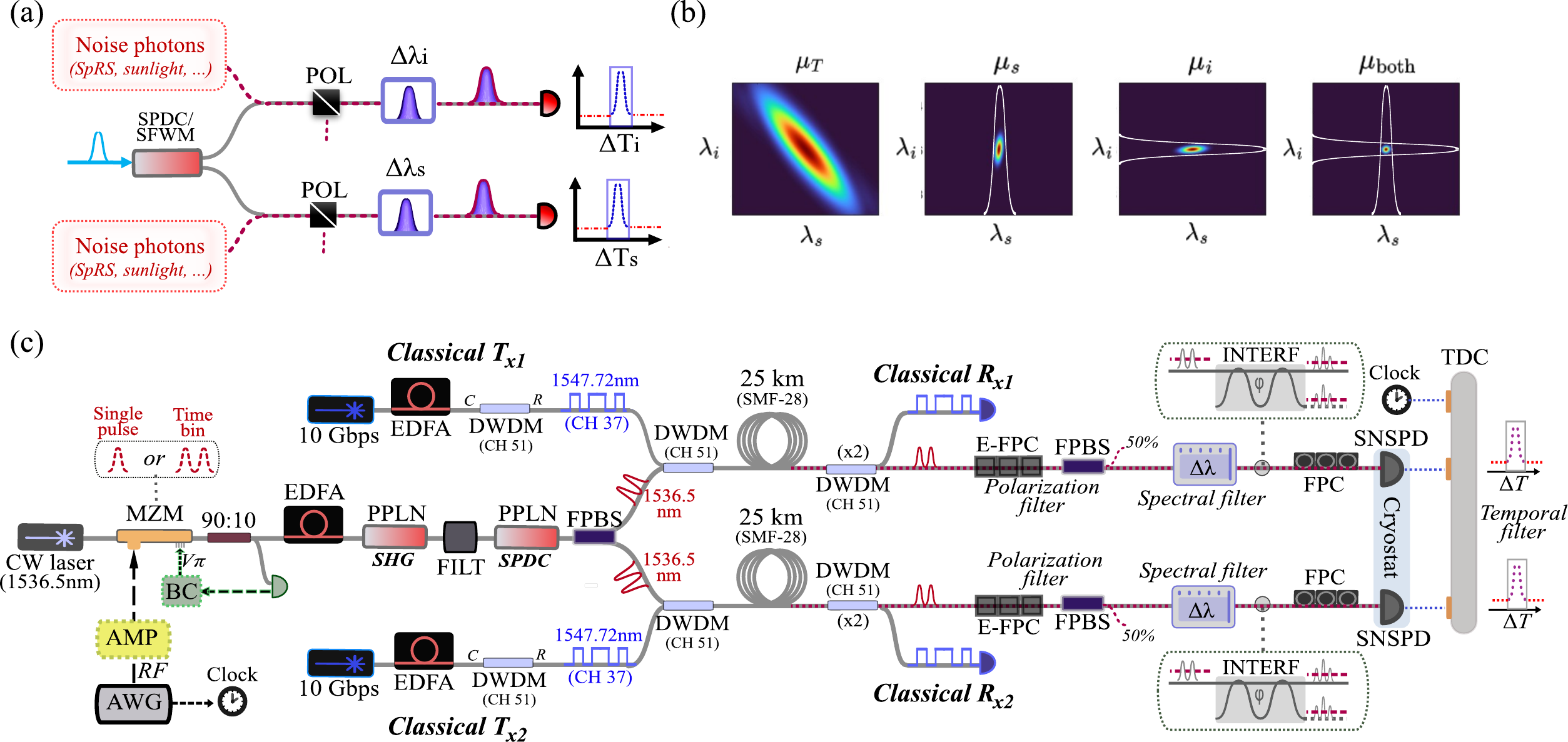}
\caption{(a) Conceptual diagram for polarization, frequency, and time filtering around photon pairs in the presence of background noise photons. (b) Effect of filtering a source's joint spectral correlations on the number of photons within the signal ($\mu_s$) and idler ($\mu_i$) filter passbands, which lead to single-photon counts, and their overlap ($\mu_{\rm both}$), which leads to true coincidence counts. When $\mu_{s/i} > \mu_{\rm both}$, not all detected photons will yield true coincidences. (c) Experimental diagram. Single-pulse or time-bin entangled photon pairs generated by type-II SPDC at degenerate wavelengths (1536.5\,nm) are distributed over two 25-km single-mode fibers to co-propagate with 10-Gbps C-band classical signals (1547.72\,nm), which generates strong spontaneous Raman scattering (SpRS) noise. We filter SpRS and SPDC photons using tunable bandwidth filters (32\,pm to 600\,pm). Polarization is filtered using a fiber polarizing beam splitter (FPBS) and an electronic fiber polarization controllers (E-FPCs) to actively correct time-dependent polarization rotations. Temporal filtering is applied in post-processing of the counts from low-jitter superconducting nanowire single-photon detectors (SNSPDs) and a time-to-digital converter (TDC). Michelson interferometers (INTERF) are inserted for time-bin experiments to perform measurements along the time-bin Bloch sphere \cite{Takesue2008_TimeBinTomography}.}
\label{fig:exp}
\end{figure*}

The deployment of advanced quantum optical technology is a widespread initiative of quantum research, which would enable quantum-enhanced encryption, sensing, and distributed quantum computation \cite{kimble_quantum_2008, wehner_quantum_2018, QNreview2022}. However, real-world environments can have high background noise that obscures the detection of single photons, limiting the fidelity or distance in which quantum information can be successfully communicated. Noise photons can originate in the transmission medium, including spontaneous Raman scattering (SpRS) when quantum and classical networks coexist in the same deployed fibers \cite{townsend_simultaneous_1997, CC_kumar_ent_1st, OvsC:2005, chapuran_optical_2009, eraerds_quantum_2010_2, dynes_ultra-high_2016_2, Mao:182,  valivarthi_measurement-device-independent_2019_2, Thomas:23, CCband_ent2, thomas_ofc_2023_2, Burenkov:23_2,  coex_ecoc23_2, 100km, CChyper2024, hyper_CC_2024_2, thomasx, Tbps_CC_narrowfilt_2024, thomasSPIE_2, CCpol_ent2025_2, Gul2025_2, OC_entdist_NYC_2025, zhang2025classical,   thomas_compare_tele2025} and ambient light in free-space quantum links \cite{freespace1, Freespace_2007_2, FreespaceFilt_2018, bouchard_achieving_2021_2, BassoBasset_et_al_2023, freespace_2024}, or in quantum devices including some photon sources \cite{kumarSFWM, noiseSFWM}, quantum frequency converters \cite{QFCnoise:18, braggQFC2024}, or optical switches and shutters \cite{premNOLM, cameronXPM, Kupchak:19}. Noise photons are most commonly rejected by frequency, time, polarization, and spatial filtering around the known modes of the photons generated at the quantum light source (Fig.~\ref{fig:exp}(a)). The design of filters for individual detectors is a well studied subject (e.g., \cite{yupingFilt2017, FreespaceFilt_2018,  coherentFilt2019, RayFilter2020, bouchard_achieving_2021_2}), including time-bandwidth limits and incoherent filtering versus coherent filtering.\\
\indent However, multiphoton quantum applications can involve more complex physics, as the design of each individual filter contributes to the overall fidelity of nonlocal multiphoton correlations. Furthermore, the physics of quantum sources must also be considered. Entangled photons are most commonly created by generating correlated photon pair states via spontaneous parametric down-conversion (SPDC) or four-wave mixing (SFWM), where filtering is also often used to increase their single-mode purity to achieve Hong-Ou-Mandel (HOM) interference \cite{HOM} between independent sources \cite{ind_photon}. This enables operations such as Bell state measurements \cite{BEll_oper}, and underlies applications such as quantum state teleportation \cite{BEll_oper}, entanglement swapping \cite{entanglement_swap}, Greenberger-Horne-Zeilinger state generation \cite{GHZ_first_1997}, photonic quantum computing \cite{opticalQC_2001, psiquantum}, and others \cite{RevModPhys.84.777, HOM_review2011}.\\
\indent Unfortunately, filtering a photon pair source's joint spectral amplitude (JSA) can degrade the ``filter heralding efficiency'' (FHE) \cite{herald1997, FHE_ultrafast2008, heraldmulti2011, FHE_simplestmodel2015, silberhorn, multistage2020, mdiQKD, highrate_timebin}, i.e., the probability that each detected signal/idler photon will be detected in coincidence with a photon at the idler/signal detector, even in a lossless environment. This issue is not resolved by simply increasing pump power, as systems with an imperfect FHE are more susceptible to degradation of fidelity from multiphoton pair emission at frequencies that never yield ``true'' coincidences between single-photon pairs \cite{heraldmulti2011, multistage2020, highrate_timebin}. Although many studies have explored engineering the source to obtain high single-mode purity without filtering \cite{FHE_elim_freq_space_Grice_Uren_Walmsley_2001, FHE_garay2007photon2007, FHE_ultrafast2008, FHE_halder2009nonclassical, FHE_theorycavity_2010, FHE_beamsilb_2011, FHE_dualpumpFWM_2013, FHE_cavity_Rielander_etal_2016, FHE_Paesani2020NearIdealPhotonSources, Liu2020_highSpectralPurityPhoton, multistage2020, FHE_Xin_dispEng_LithNio_2022, psiquantum}, noisy environments still require narrow external filtering of one or more photons.\\% In many cases, filtering to improve noise rejection and single-mode purity comes at the expense of coincidence rates and single-photon purity. 
\indent If the reduction in coincidence rates exceeds the rejection of noise, improvements in fidelity may be hindered compared to systems with ideal FHE. This further influences the design of multiphoton applications when filtering also affects the single-mode purity.\\
\indent In this paper, we explore these concepts in the context of filtering SpRS noise photons generated when classical communication signals propagate in the same long-distance fibers as entangled photons, which would enable large-scale quantum networking within the existing fiber infrastructure and the ability to use classical signals to perform quantum node synchronization, routing, and error mitigation. Although ``quantum-classical coexistence'' in fiber has been well explored using weak coherent state (WCS) sources and a single detector \cite{townsend_simultaneous_1997, OvsC:2005, chapuran_optical_2009, eraerds_quantum_2010_2, dynes_ultra-high_2016_2, Mao:182, Tbps_CC_narrowfilt_2024}, recent studies have focused on identifying the challenges facing more complex applications using entangled photons \cite{CC_kumar_ent_1st, Thomas:23, CCband_ent2, coex_ecoc23_2, 100km, CChyper2024, hyper_CC_2024_2, CCpol_ent2025_2, OC_entdist_NYC_2025, Gul2025_2} and quantum teleportation \cite{thomasx, thomasSPIE_2, thomas_compare_tele2025}.\\ 
\indent Since the SpRS spectrum has a broad bandwidth spanning the full 1260\,nm to 1625\,nm range of telecom wavelengths (>53\,THz) \cite{thomas_ofc_2023_2, Burenkov:23_2, Thomas:23}, filtering is critical to maximize quantum and classical capabilities. However, allocating quantum and classical signals to different telecom bands suppresses SpRS due to the multimode Raman gain spectrum and the phonon population, both highly dependent on the frequency separation of the signals \cite{OvsC:2005, chapuran_optical_2009, eraerds_quantum_2010_2, valivarthi_measurement-device-independent_2019_2, thomas_ofc_2023_2, Burenkov:23_2, Thomas:23}. For example, using O-band (1260\,nm-1360\,nm) quantum sources and narrow filtering can enable applications based on WCSs \cite{Mao:182}, entangled photons \cite{Thomas:23}, and quantum teleportation \cite{thomasx, thomasSPIE_2} to operate alongside C-band (1530\,nm-1565\,nm) classical power on the order of 20\,dBm. However, the O-band sacrifices transmission loss for noise suppression ($\sim$0.33\,dB/km in standard fiber), whereas quantum-classical coexistence in the C-band using dense-wavelength division multiplexing (DWDM) enables both network paradigms to leverage the lowest-loss telecom wavelengths ($\sim$\,0.18\,dB/km). The main challenge lies in the significantly higher Raman gain and phonon population, which makes it difficult to engineer discrete-variable systems that can tolerate \mbox{$>$\,-10\,dBm} of C-band power \cite{CC_kumar_ent_1st, chapuran_optical_2009, eraerds_quantum_2010_2, dynes_ultra-high_2016_2, valivarthi_measurement-device-independent_2019_2, CCband_ent2, coex_ecoc23_2, hyper_CC_2024_2, Tbps_CC_narrowfilt_2024, thomasSPIE_2, CCpol_ent2025_2}. Relatively high classical rates have been achieved by attenuating classical power, but much higher power is common in the current fiber infrastructure. C-band quantum networking alongside mW-level C-band classical communications has thus been a long-standing goal.\\
\indent To experimentally investigate this, we distribute either single-pulse or time-bin entangled photon pairs at degenerate wavelengths (1536.5\,nm) over 50\,km of standard single-mode fiber, where each photon co-propagates over 25-km fibers with a 10-Gbps C-band (1547.72\,nm) classical data signal (Fig.~\ref{fig:exp}(c)). We first study the filtering of correlated photon pairs and show that imperfect FHE has a negative effect on performance. Then, we apply narrow spectrotemporal and polarization filtering to distribute pulsed time-bin entangled photons in the C-band alongside mW-level C-band power, which is an order of magnitude higher than comparable C-band/C-band experiments using entangled photons \cite{CC_kumar_ent_1st, CCband_ent2, coex_ecoc23_2, CChyper2024, hyper_CC_2024_2, CCpol_ent2025_2}. We then quantify how optimizing FHE, loss, and pump power could improve tolerance to classical power or noise.\\
\indent By simulating various system designs, we find that fidelity depends on the pump pulse width, filter bandwidth and shape, pump power, and phase matching. For example, flat-top filters are shown to achieve higher noise tolerance and purity compared to Gaussian filters. Furthermore, we find that the duration of the pump pulse can notably impact performance since the FHE is lessened at wider filter bandwidths, where optimizing the entanglement fidelity can come at the expense of single-mode purity. We consider the impact of multipair emission for each case, showing that the trade-off between the single-detector signal-to-noise ratio (SNR) and degradation due to multipairs is more nuanced for systems with an unideal FHE.\\
\indent This work highlights further considerations for the design of sources and receivers beyond single-detector filtering and applies to many scenarios beyond quantum-classical integration, such as noise in free-space links and quantum devices. Our results and methods have implications for many experiments using filtering to either reduce noise or increase single-mode purity and for high-repetition-rate sources requiring short pulses and wider filters. Furthermore, due to the 1536.5-nm wavelength of both entangled photons, this study provides insight into the challenges of deploying quantum memories based on erbium-ions \cite{Er2010, ErQMtele2025} into noisy fiber networks carrying classical signals. The ability to co-propagate high classical power in nearby C-band DWDM channels shows promise for scalable low-loss entanglement-based networks in existing infrastructure. More broadly, our thorough investigation of the role of multipair emission given imperfect FHE in noise-free environments provides useful methods for designing general multiphoton applications. \section{Filtering Noise and Correlated Photon Pairs}\label{section1}
\begin{figure}[!t]
\centering
\includegraphics[width=1\linewidth]{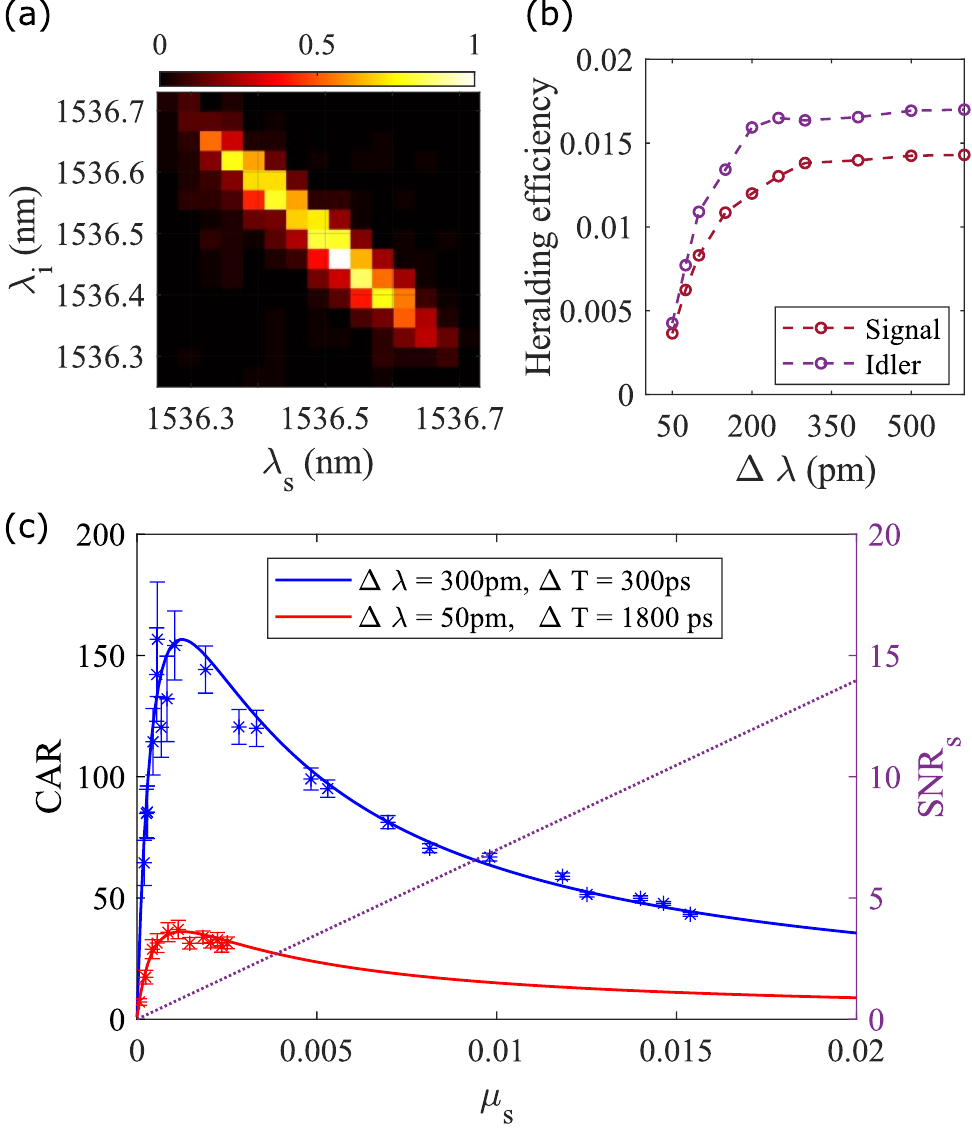}
\caption{(a) Measured JSI of the type-II SPDC photon pair source. (b) Heralding efficiencies as a function of filter bandwidth $\Delta \lambda$. (c) CAR as a function of $\mu_s$ for \mbox{$\Delta \lambda \Delta T =300\,\rm pm \times 300 \rm ps$} or \mbox{$\Delta \lambda \Delta T =50\,\rm pm \times 1800\,\rm ps$}, which have different FHEs but the same background count probability since $\Delta \lambda \Delta T$ is fixed. The classical power is adjusted to receive $P_R = -27$\,dBm after each 25-km fiber. The solid lines show the prediction based on our model. Even when each case has the same single-detector SNR (purple), the two-fold CAR for $\Delta \lambda=50\,$pm is notably lower compared to $\Delta \lambda=300\,$pm} 
\label{300_50}
\end{figure}

Our experimental design is shown in Fig.~\ref{fig:exp}(c). We generate C-band photon pairs (1536.5\,nm) using type-II SPDC and multiplex them into fibers using DWDMs (ITU channel 37) to co-propagate with 10-Gbps 1547.72-nm (ITU channel 51) classical data signals over 25\,km of standard single-mode fiber (Corning SMF-28).\\
\indent Frequency-domain filtering is studied using tunable-bandwidth optical filters (EXFO {XTA-50/U}) with a full width at half maximum (FWHM) bandwidth range of 32\,pm to 600\,pm. Temporal filtering is applied by defining a detection time window around the pulse arrival times during post-processing of the count time-stamps from our superconducting nanowire single-photon detectors (SNSPD, Quantum Opus) and time-to-digital converter (TDC, Swabian Timetagger Ultra). Each has a root mean squared timing jitter of 45\,ps and 8\,ps, respectively, enabling tight filtering. Each SNSPD has a dead time of <25\,ns, an efficiency of >90\%, and dark count rate of $<$100\,counts/s, which means that the vast majority of noise comes from SpRS.\\
\indent Polarization filtering is applied by placing fiber polarizing beam splitters (FPBSs, {Thorlabs}) at the end of each fiber. Although Raman gain has some polarization dependence \cite{Raman_pol_dependent}, wavelength-dependent birefringence and polarization mode-dispersion over long-distance fiber causes SpRS photons to be approximately unpolarized \cite{polTomo2023, Thomas:23, thomasx}, where the FPBSs reject $\sim$50\% of the noise exiting the fiber. To correct for time-dependent rotations of polarization in the fibers due to environmental disturbances, we implement active alignment to the polarizer by maximizing single-photon counts using electronic fiber polarization controllers (EFPCs, {Luna Inc.}).\\
\indent Photon pairs are generated by type-II SPDC in a periodically poled lithium niobate waveguide (PPLN-WG, {HC Photonics}). Our source is pumped at a repetition rate of 200\,MHz by intensity modulating a 1536.5-nm continuous wave (CW) laser (Pure Photonics {PPCL550}) using a 20-GHz bandwidth lithium niobate Mach-Zehnder modulator (MZM, Exail {MXER-LN-20}). The MZM is driven by amplified radio frequency (RF) pulses originating from an arbitrary waveform generator (AWG, Tektronix {AWG70002}), which produces pulses with a temporal width of approximately 50\,ps (FWHM). A 90:10 splitter is used to tap the pump light as input to a MZM bias controller (BC, Exail) to stabilize the extinction ratio of the pulses. The pump pulses are then amplified using an erbium-doped fiber amplifier (EDFA, Pritel {PMFA-20}) and injected into a PPLN-WG to undergo second harmonic generation (SHG), producing pulses centered on 768.25\,nm. After filtering out the 1536.5-nm pump, the 768.25-nm pulses drive type-II SPDC in a second PPLN-WG (HC Photonics). SPDC generates a two-mode squeezed vacuum state \cite{TMSV88}, which is a probabilistic source of correlated photon number states in the Fock state basis with an imperfect but high single-photon fidelity in the low gain regime. The photons in each mode (signal and idler) are orthogonally polarized and separated using a FPBS. Each photon is subsequently multiplexed with the 10-Gbps C-band classical light using the pass and reflect ports of a 100-GHz DWDM (ITU channel 37) to co-propagate over 25-km SMF-28 fiber spools.\\
\indent The measured joint spectral intensity (JSI) of our source is shown in Fig.~\ref{300_50}(a), which is pre-filtered by a 100-GHz DWDM and measured to a resolution of 32\,pm. Fig.~\ref{300_50}(b) shows the measured HE as we decrease the bandwidth of the filters for each photon with the classical sources off (negligible noise), which is defined as $\eta'_{s/i} =  C /S_{i/s}$, where $C$ and $S_j$ are the coincidence and singles count rates, respectively. It is clear in Fig.~\ref{300_50}(b) that $\eta'_{j}$ is notably impacted as we decrease the filter bandwidth.\\
\indent Each of the classical data signals are generated by small-form programmable transceivers (FS {SFP-10G-40}) that require a minimum received power of \mbox{$P_R = -18$\,dBm} to communicate with a sufficiently low bit error rate. The quantum and classical light experienced approximately 5-dB transmission loss over the 25-km fibers. The minimum launch power to operate each classical link was $P_0 = -12.5$\,dBm when accounting for a 0.5-dB loss in the demultiplexing DWDMs. To test the limits of classical power and background noise, the classical signals are amplified by an EDFA before multiplexing with the quantum signal. We filter the classical signals before transmission using DWDMs to prevent 1536.5-nm amplified spontaneous emission from the laser and EDFA from entering the 25-km fibers. At the quantum receivers, we cascaded multiple DWDMs until negligible noise counts were observed when directly injecting 1547.72-nm light within the tested power range. The above ensures that the only appreciable noise comes from SpRS along the fibers and is within the passband of the 1536.5-nm filters.\\
\indent To model how filtering affects single- and two-fold detection rates, we note that an imperfect FHE implies that the mean number of photons that are generated at frequencies within the signal ($\mu_s$), idler ($\mu_i$), and overlap ($\mu_{\rm both}$) of the filter passbands are not equal. This is represented pictorially in Fig.~\ref{fig:exp}(b). When filtering the JSA leads to $\mu_{j} > \mu_{\rm both}$, the FHE is $\delta_{s/i} = \mu_{\rm both}/\mu_{i/s}< 1$. This means that even when the true signal/idler channel efficiencies are ideal ($\eta_s = \eta_i = 1$), not all detected single photons result in two-fold coincidence events between single-photon pairs. However, SPDC and SFWM sources also emit multiphoton pairs, which leads to accidental coincidences or imperfect fidelity for entangled pair sources \cite{takesue_effects_2010_2, Takeoka_2015}.\\
\indent For an unfiltered JSA represented by $f(\omega_s, \omega_i)$ with $\iint d \omega_s d \omega_i\left|f\left(\omega_s, \omega_i\right)\right|^2 \equiv 1$, we define $\mu_T$ as the mean number of photon pairs per pulse generated across the unfiltered JSI. When filters with a normalized filter transmission function $g_j\left(\omega_j\right)$ are applied to each photon, we define the mean number of photons within the passbands of the signal/idler filters individually as (see Appendix \ref{appendix_model})
\begin{equation}\label{muj}
     \mu_j=\mu_{T} \iint d \omega_s d \omega_i\left|f\left(\omega_s, \omega_i\right) g_j\left(\omega_j\right)\right|^2,
\end{equation}
\noindent which quantifies the photons that can lead to signal ($\mu_s$) or idler ($\mu_i$) single-detector counts, and the number of photons within both filters
\begin{equation}\label{coinc}
\mu_{\mathrm{both}}=\mu_T \iint d \omega_s d \omega_i\left|f\left(\omega_s, \omega_i\right) g_s\left(\omega_s\right) g_i\left(\omega_i\right)\right|^2,
\end{equation}
\noindent which quantifies the number of photons that can lead to ``true'' coincidence counts. However, multipairs populate the entire JSI, which becomes more problematic when $\mu_j > \mu_{\rm both}$ because multipairs are detected at frequencies that do not contribute to single-pair coincidences.\\
\indent To model our experiment in the low-gain regime, we modify the models in refs.~\cite{Takesue2006_TimeBinDistribution,  takesue_effects_2010_2, Takeoka_2015} by substituting \mbox{$\mu \rightarrow \mu_{s}, \mu_i$} (eq.~\ref{muj}) and including a filtered noise term to obtain the single-detector count probability per gate (see Appendix \ref{appendix_model})
\begin{equation}
S_j=\mu_j \eta_j+\left(\eta_{r_j} \alpha_{\mathrm{pol}} \mathcal{R}_j \Delta \lambda_j+d_j\right) \Delta T_j
\end{equation}
\noindent and replace $\mu \rightarrow \mu_{\rm both}$ (eq.~\ref{coinc}) to obtain the two-fold coincidence probability
\begin{equation}
    C \approx \mu_{\rm b o t h}  \eta_s \eta_i + S_s S_i.
\end{equation}
\indent In the above, we represent an arbitrary noise photon source to be filtered by the parameter $\mathcal{R}_{j}$, which has units of $\mathrm{nm}^{-1} \mathrm{s}^{-1}$. Since noise photons are typically filtered to a narrow bandwidth $(\Delta \lambda <1\,\rm nm)$, we assume that the noise is approximately constant across the passbands of the frequency- and time-domain filters, which is an accurate approximation for broadband SpRS generated by an uncorrelated pump in long-distance fiber. $\Delta \lambda_j$ is the FWHM bandwidth of the spectral filter and $d_j$ is the detector's dark count rate per second. $\Delta T_j$ is the detection time window, which temporally filters both background noise and dark counts. We assume at present that $\Delta T_j$ is greater than the temporal width of the photon pairs (see Section~\ref{conlusion_and_outlook}). The channel efficiency for each photon can be broken down as $\eta_j = \eta_{c_j} \eta_{\rm ch_j} \eta_{r_j} \cos^2(\theta_{pol_j})$ where $\eta_c$, $\eta_{\rm ch}$, and $\eta_r$ account for the loss at the source, transmission channel, and receiver, respectively. Maximizing $\eta_{c_j}$ is critical for our experiment, since minimizing the loss acting solely on the source photons will maximize the SNR in each detector. $\alpha_{\rm pol}$ is a polarization filtering parameter that reduces the noise count rate, which is equal to 1/2 for unpolarized noise. The term $\cos^2(\theta_{\rm pol})$ represents the transmission of source photons through the polarization filter.\\
\indent The accidental coincidence probability is modeled as \mbox{$A = S_s S_i$}, which is the same as refs.~\cite{Takesue2006_TimeBinDistribution,  takesue_effects_2010_2, Takeoka_2015} but with the new definitions for $\mu_s$ and $\mu_i$. We can then define the coincidence-to-accidental ratio (CAR) to be \mbox{$\text{CAR} = C/A$}, which is the two-photon analogue of the single-photon SNR. When the background noise is negligible, the HE including channel loss is defined as \mbox{$\eta_{s/i}' = C/S_{i/s} \approx \delta_{s/i} \eta_{s/i}$}, and \mbox{$\delta_{s/i} = \mu_{\rm both}/\mu_{i/s}$} is the FHE. When considering imperfect FHE of both photons, it is useful to use the "pair-symmetric heralding efficiency" (PSHE), which is defined as $\delta_{\rm PS} \equiv \sqrt{\delta_s \delta_i}$ \cite{silberhorn}.

%\subsection{Results}

To confirm that imperfect FHE affects performance and how this differs from single-detector filtering, we first compare two scenarios with unequal FHEs but the same single-detector SNR. To do this, we evaluate two cases of spectrotemporal filtering in which one has a wider spectral bandwidth but narrower temporal filtering ($\Delta \lambda_1 =300$\,pm, $\Delta T_1 =300$\,ps) and one which has a narrower bandwidth but wider temporal filtering ($\Delta \lambda_2 =50$\,pm, $\Delta T_2 =1800$\,ps), such that \mbox{${\Delta \lambda_1 \Delta T_1} = {\Delta \lambda_2 \Delta T_2}$}. For a single detector, the SNR is \mbox{$\operatorname{SNR}_j={\mu_j \eta_j}/(\eta_r \alpha_{\rm pol}{\mathcal{R}_j \Delta \lambda_j \Delta T_j})$}, assuming dark counts are negligible. Since $\operatorname{SNR}\propto 1/{\Delta \lambda \Delta T}$, two different receivers with the same value for the product ${\Delta \lambda \Delta T}$ have an equivalent SNR whenever $\mu_j$ is equal. This also holds for the CAR when $\delta_{\rm PS} = 1$. However, we predict that even if the SNR is the same, the CAR of a system with the lowest $\delta_{\rm PS}$ will be more susceptible to noise. For our system, Fig.~\ref{300_50}(b) shows that the FHE when $\Delta \lambda = 50$\,pm is significantly lower compared to $\Delta \lambda = 300$\,pm, which implies that 50-pm filtering should have a lower CAR.\\
\indent We first measure the CAR as a function of $\mu_s$ for 300-pm/300-ps filtering versus 50-pm/1800-ps filtering. For this test, the classical sources are attenuated to obtain a received power of \mbox{$P_R = -27$\,dBm} at the output of each 25-km fiber spool. This gave a noise count rate within $\Delta \lambda_{} \Delta T_{}$ of $5180.4 \text{ counts/s}$ and $5050.5 \text{ counts/s}$ for signal and idler detectors, respectively.\\%We note that the SNSPD dark-count rates were low enough compared to SpRS to have a negligible impact for these measurements.\\
\indent The results are shown in Fig.~\ref{300_50}(c). For 300-pm/300-ps filtering we measured a maximum CAR of $\approx$\,155, while 50-pm/1800-ps obtains a much lower maximum of $\approx$\,40. This clear difference between the two cases unequivocally confirms that filtering the JSA has a notable impact. Since the CAR differed even when the SNR was the same, this also confirms a clear difference between filtering photon pairs compared to filtering single photons.\\
\indent We also observe some consequences of balancing the SNR and degradation due to multipair emission during SPDC. Interestingly, the CAR can reach a high value even when the SNR is insufficiently low. We attribute this to the temporal and photon-number correlations for SPDC/SFWM, which are only benefited from in coincidence experiments. However, the CAR begins to decrease as the SNR increases. The maximum CAR occurs near $\mu_s \sim 10^{-3}$, but decreases as $\mu_s \rightarrow 0$ and $\mu_s \rightarrow \infty$. When the source's pump power is low, multipair emission is optimized at the expense of a lower SNR. When the SNR is increased, multiphoton emission begins to degrade CAR and eventually becomes a dominant source of "noise." The maximum $\mu_s$ for 50-pm filtering in Fig.~\ref{300_50}(c) was more limited than 300-pm due to the narrower bandwidth, which requires higher pump power to produce the same number of photons. However, our results show that simply increasing the source's pump power does not compensate for this when $\delta_{\rm PS} < 1$, which is partially due to the fact that each case is not affected equally by accidental coincidences due to multipair emission (see Section~\ref{simulation}).\\
\indent Instead of fixing the noise count probability, we now fix $\Delta T=300$\,ps and tune $\Delta \lambda$. We also increase the launch power of the classical signals to $P_0 = -12.5$\,dBm so that $P_R$ was greater than the minimum receiver sensitivity of the 10-Gbps transceivers. This results in the SpRS noise counts of $ 1477.4$\,counts/pm/s and $1040.1$\,counts/pm/s for signal and idler detectors, respectively, where the difference is mainly attributed to unequal detection efficiencies. For this measurement, we fix the SPDC pump power to its maximum setting. We note that the SPDC pump power does not increase as $\Delta 
\lambda$ is varied, which does not create a perfect comparison, since both the number of SPDC and SpRS photons change with $\Delta \lambda$. Narrowing $\Delta \lambda$ therefore decreases multipair emission and noise count rates, but also does not improve the SNR. However, it allows us to observe some of the consequences of narrow filtering when $\Delta T$ is fixed, in contrast to the previous measurement. In Section~\ref{simulation}, we simulate the case where we increase the SPDC pump power as $\Delta \lambda$ is decreased to maintain a fixed source intensity.\\ \begin{figure}[!t]
\centering
\includegraphics[width=1\linewidth]{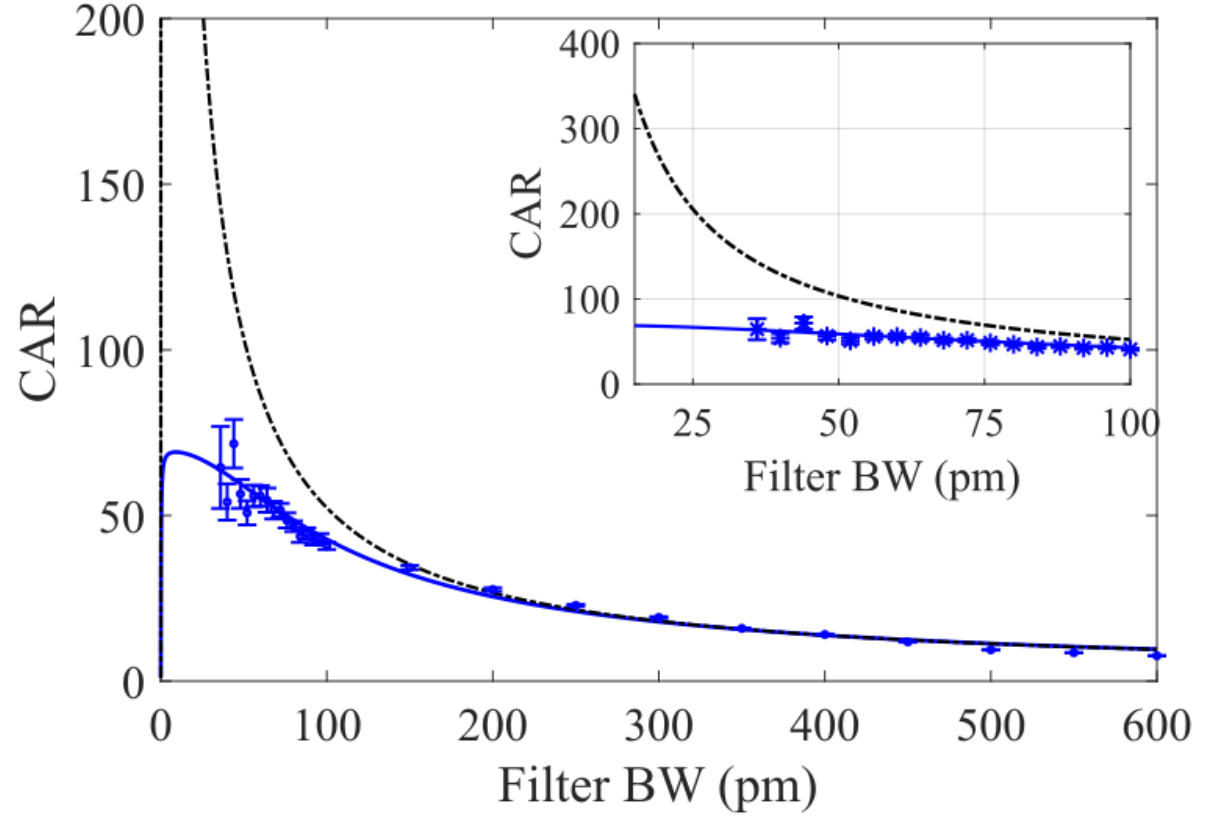}
\caption{CAR as a function of the spectral filter bandwidth (BW) at the signal/idler receivers $\Delta \lambda_{s,i}$ for a fixed SPDC pump power whilst each C-band classical source transmits $P_0 = -12.5$\,dBm into each 25-km/25-km fibers. The black line shows the simulation for an ideal system, and the blue line shows the results accounting for $\delta_{j}(\Delta \lambda_{j})<1$. The inset shows the results within the range of $\Delta \lambda_{j}$=\,20\,pm to 100\,pm.}
\label{CARbw}
\end{figure}
\indent The results are shown in Fig.~\ref{CARbw}. We observe that the measured CAR (blue) begins to diverge significantly from the ideal model assuming a perfect FHE (black dashed line) for $\Delta \lambda < 200$\,pm, which coincides with the trend in the change in HE in Fig.~\ref{300_50}(b). For example, a \mbox{$\mathrm{CAR_{ideal}}(32\,\mathrm{pm})= 155$} is predicted when $\delta_{\rm PS} =1$, while the measured result is \mbox{$\mathrm{CAR_{meas}}(32\,\mathrm{pm})= 71.3 \pm 5.0$}. Since a decrease in FHE usually coincides with an increase in single-mode purity \cite{silberhorn}, this shows a potential trade-off between minimizing the impact of noise photons and maximizing the raw fidelity of applications requiring high-visibility HOM interference. We explore the consequences of this in more detail in Section~\ref{simulation}.
\section{C-band time-bin entanglement coexisting with C-band classical communications}
\begin{figure*}[!t]
\centering
%\onecolumn
\includegraphics[width=\textwidth]
{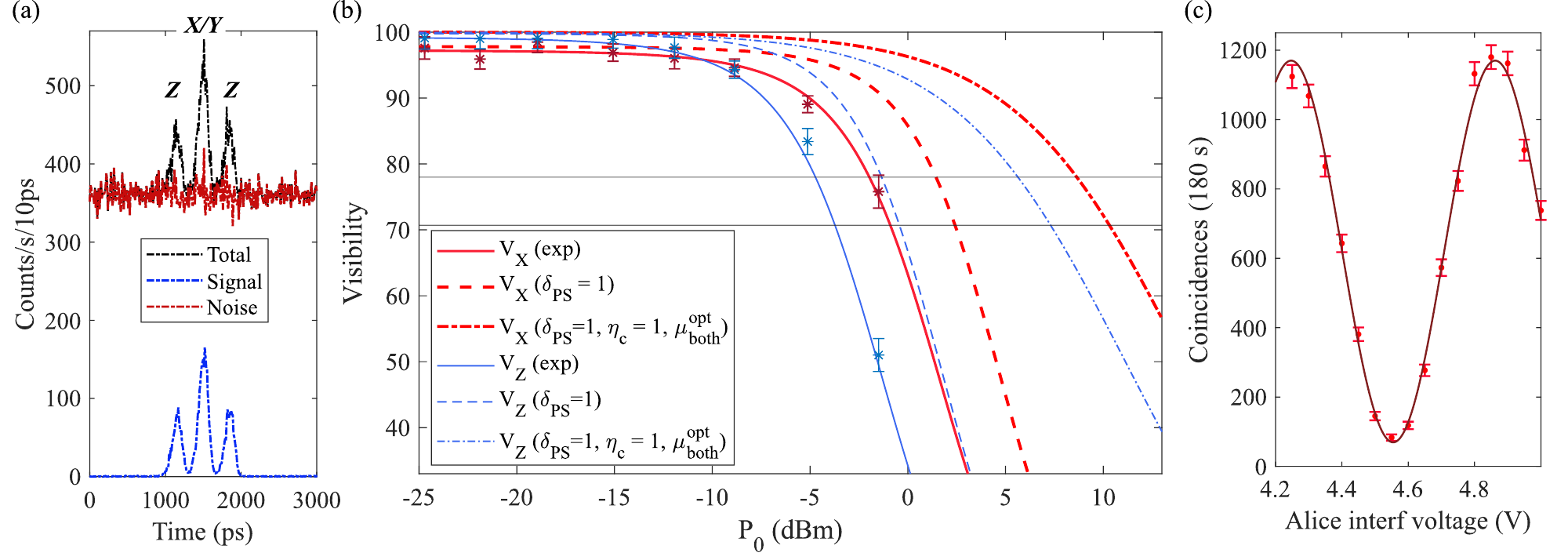}
\caption{Results for distributing 1536.5-nm time-bin entangled photons over 25-km/25-km fibers with a co-propagating 10-Gbps C-band (1547.72\,nm) classical data signal in each fiber. The filtering is set to $\Delta \lambda=50$\,pm and $\Delta T =200$\,ps for each photon. (a) Histogram of single counts relative to the clock for signal time bin photons, SpRS noise, and total counts at the output of Alice's interferometer when the classical launch power is $P_0=-4.8$\,dBm. (b) Entanglement visibility in the $X$ and $Z$ bases versus $P_0$. The color-coded solid lines show the predictions given our experimental parameters. The dashed lines are simulations when optimizing parameters such as ideal PSHE ($\delta_{\rm PS} = 1$), source loss ($\eta_{c} = 1$), or optimization of the entangled pair source's pump power at each $P_0$. The black horizontal lines show the limits for QKD and verifying nonlocality. (c) Two-photon interference fringe when $P_0=-4.8$\,dBm, where $V_X = 89.6 \pm 1.6$\%.}
\label{timebin}
\end{figure*}
\indent We now investigate filtering for a concrete application, namely entanglement-based networking in long-distance fibers carrying classical communications. Although polarization entanglement \cite{CC_kumar_ent_1st, Thomas:23, coex_ecoc23_2, CCpol_ent2025_2, thomasx} and time-energy entanglement \cite{CCband_ent2} have been studied, to our knowledge the impact of SpRS on pulsed time-bin entanglement has not yet been investigated. Pulsed time-bin encoding fundamentally requires a spectrally broadband pump, especially at higher repetition rates, meaning that narrow filtering can reasonably impact the FHE of many systems.\\
\indent To generate entangled photons, we use the AWG to drive the MZM with two equal amplitude RF pulses with a time separation of 347\,ps at the same 200-MHz repetition rate. Since our 1536.5-nm CW laser has a $\sim$$30 \,\mu$s coherence time, the pump is in a coherent superposition of being in 'early' and 'late' time bins: \mbox{$|\psi\rangle_p = \frac{1}{\sqrt{2}}(| E_p\rangle +| L_p\rangle)$}. After amplification by the EDFA and SHG at the first PPLN-WG, the time-bin encoded 768.25-nm pump probabilistically generates single-photon pairs in the \mbox{$\left|\Phi^{+}\right\rangle=\frac{1}{\sqrt{2}}\left(\left|E_s, E_i\right\rangle+\left|L_s, L_i\right\rangle\right)$} Bell state \cite{Brendel1999}. Michelson 1-bit delay interferometers are then inserted at each receiver to perform projective measurements on the time-bin Bloch sphere \cite{Takesue2008_TimeBinTomography}.\\
\indent In this experiment, we fix the filtering to \mbox{$\Delta \lambda_{j} = 50$\,pm} (6.5 GHz) and $\Delta T_{j} = 200$\,ps for both photons. This means that each filter is close to time-bandwidth limited and also gives a ratio between the temporal width of the filtered photon and pump pulse of $\tau_{ph}/\tau_p \approx 1.5$ (see Appendix\,\ref{appendix_section1}), which implies that both photons are forward compatible with HOM interference-based applications. However, the FHE is notably impacted, which we measured to be approximately $\delta_s = 0.23$ and $\delta_i = 0.21$ for signal and idler photons, respectively. The heralding efficiencies including loss were $\eta_s' = 0.0012$ and $ \eta_i'=0.0014$, which is lower than previous measurements due to the interferometers. At the expense of the tightly filtering background noise and improved spectral purity, this is expected to hinder our results.\\
\indent Fig.~\ref{timebin} shows the results of distributing the 1536.5-nm time-bin entangled pairs over 25-km spools to co-propagate with the 10-Gbps C-band classical signals. Fig.~\ref{timebin}(a) shows a histogram of the signal detector's single-photon counts per 10-ps bin width relative to the arrival of our clock at the TDC when \mbox{$P_0 = -4.8\,\rm dBm$}. We plot the source and noise counts individually to distinguish how the interferometer affects the arrival time of time-bin versus SpRS photons. For the time-bin photons, we observe pulses within three distinct time slots, with the middle bin ($X/Y$ bases) having double the counts as the two side bins ($Z$ basis) due to photons combining after taking the short and long paths through the interferometer. However, the noise has an equal intensity across all time, indicating that SpRS from the classical source is generated randomly in time relative to the source. The noise photons are $1/2$ the input resulting from randomly exiting the second beam splitter of the interferometer. The SpRS noise count rates increased linearly as a function of classical power with 145793.8 counts/s/mW and 158694.0 counts/s/mW in the signal and idler detectors, respectively.\\
\indent Fig.~\ref{timebin}(b) shows how the quality of the time-bin entangled photons changes as we increase $P_0$ in each fiber. Nonlocal two-photon interference (TPI) fringes are observed in coincidence counts between the middle bins when varying Alice's interferometer phase ($\phi_{A}$) while fixing Bob's ($\phi_{B}$) (Fig.~\ref{timebin}(c)), which cycles through projective measurements along the equator of the Bloch sphere \cite{Takesue2008_TimeBinTomography}.
To quantify the degree of entanglement, we calculate \mbox{$V=(R_{ll} - R_{lk})/(R_{ll} + R_{lk})$} for each measurement, where $R_{ll}$ and $R_{lk}$ are the maximum and minimum coincidences, respectively. When the classical sources are off ($P_0 = 0\,\rm mW$), we measure $V_Z =  99.1 \pm 0.9\%$ and $V_X = 97.1 \pm 1.6\%$ in the $Z$ and $X$ bases, respectively, where the $X$ basis is more degraded due to imperfections in the interferometers. Both bases are less than unity because of multipair emission in orthogonal bases. For these measurements we kept a fixed SPDC pump power, which we estimated yielded approximately $\mu_s = 1.0 \times 10^{-3}$, \mbox{$\mu_i = 1.1 \times 10^{-3}$}, and $\mu_{\rm both} = 2.5 \times 10^{-4}$.\\ 
\indent As we increase $P_0$, Fig.~\ref{timebin}(b) shows that our system is able to tolerate relatively high C-band classical power for a C-band/C-band quantum/classical wavelength allocation. In the $X$ basis, we predict that $V> 78\%$ could be maintained at $P_0=-1.7$\,dBm and $V>70.7\%$ up to $P_0=-0.4$\,dBm, which are the limits for QKD \cite{Jin:19} and verifying nonlocality \cite{CSHS}, respectively. However, $V_Z$ could maintain these values up to $P_0=-4.6$\,dBm and $P_0=-3.5$\,dBm, respectively. We attribute this to the unequal SNR of each bin in the output of the interferometers, as can be seen in the single-photon measurement in Fig.~\ref{timebin}(a). This could be mitigated by measuring the $Z$ basis in an alternative way, such as taking out the interferometer (at the expense of measuring $X/Y$) or using a switch or splitter to bypass the interferometer for $Z$-basis measurements (see Appendix~\ref{appendix_timepol}).\\
\indent To investigate how imperfect FHE limited our results and how more optimal systems could be designed, we model time-bin entanglement generation, filtering, and measurement. We assume that the mean number of photons generated per bin is \mbox{$\mu_j \equiv \mu_j^E = \mu_j^L$}, which is required for the state to be maximally entangled. Building off of the models in refs.\,\cite{takesue_effects_2010_2, kim2022quantum}, we model the coincidence and singles count probability for the $X/Y$ bases as (see Appendix~\ref{appendix_timepol}):
\begin{equation}
C^{xy}(\phi)=\frac{1}{4} \mu_{\mathrm{both}} \eta_s \eta_i\left(\frac{1}{2}+\frac{1}{2} V_{\mathrm{int}} \cos \left(\phi\right)\right)+S_s^{xy} S_i^{xy}
\end{equation}
\begin{equation}
	S_{j}^{xy} = \frac{1}{2}\mu_{j} \eta_j + \left(\frac{1}{2}\eta_{r_j}\alpha_{pol}\mathcal{R}_j \Delta \lambda_j+d_j\right) \Delta T_j,
\end{equation}
\noindent where $\phi \equiv \phi_A+\phi_B$ and $V_{\rm int}$ accounts for the visibility of the interferometer. The background noise term is 1/2 the input due to splitting and recombining at the two interferometer beam splitters. The 1/4 reduction in the coincidence probability is due to how the time-bin photons are distributed across the three bins of the two interferometer outputs \cite{takesue_effects_2010_2}. The $Z$ basis obtains a 1/16 reduction in coincidence rates, leading to a higher susceptibility to incident noise compared to the $X/Y$ bases when using only the output ports of the interferometers. In Appendix~\ref{appendix_timepol}, we investigate different measurement schemes and compare time-bin to polarization entanglement distribution in noisy environments.\\
\indent In Fig.~\ref{timebin}(b), we plot the predictions according to our experimental parameters (solid lines) and investigate the impact of optimizing different parameters (dashed lines). We find that our source was not generating enough photons to obtain the highest possible visibility by overcoming the high noise rates. We predict that the maximum classical power where we could still maintain $V_X>70.7\%$ is $P_0=2.11$\,dBm. We also simulate our system assuming the same pump power but with perfect FHE to quantify how much the FHE impacted our measurements, which we predict could have tolerated up to $P_0 \approx 3$\,dBm. Another parameter that can be optimized is the loss of photons at the source, since the SpRS photons do not experience this loss. In our system, photons can be lost due to coupling from the PPLN-WG into fiber and insertion loss in the FPBS and DWDMs, which totaled approximately 2.1\,dB. Assuming ideal PSHE, source loss, and optimizing the pair generation probability for each $P_0$, the simulation in Fig.~\ref{timebin}(b) predicts that $V_{X/Y}>70.7\%$ can be maintained with $P_0 \approx$ 9.8\,dBm in each fiber.\\
\indent Even without further optimization, filtering in multiple degrees of freedom has enabled us to achieve notably higher C-band power compared to past C-band entangled photon pair experiments. However, the above simulations can guide the design of future quantum networks to have an even higher noise tolerance.
\section{Considerations in the design of optimal sources and filters}\label{simulation}
\begin{figure}[!t]
\centering
\includegraphics[width=1\linewidth]{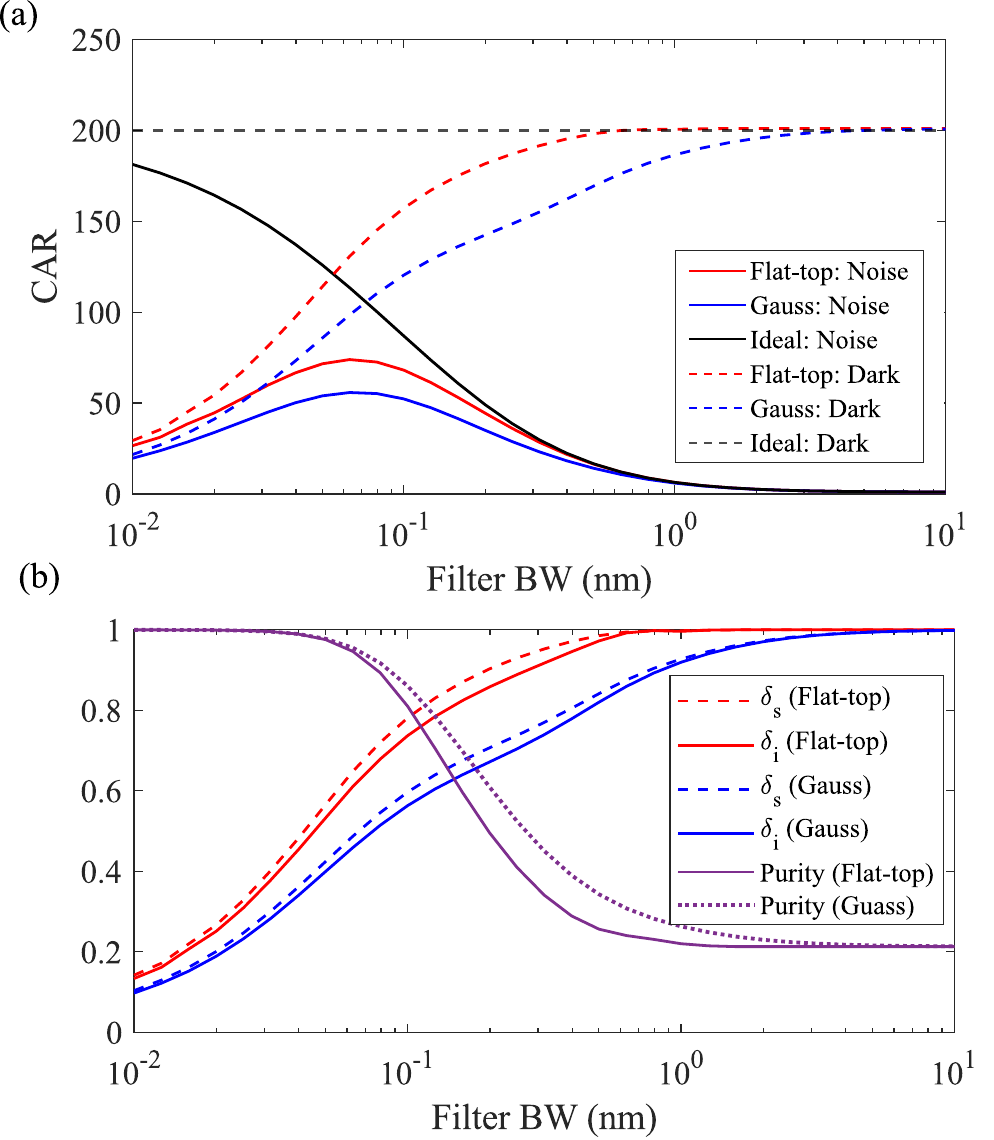}
\caption{(a) CAR as a function of filter bandwidth for either Gaussian or flat-top filter shapes when $P_0=-12.5$\,dBm. The dashed lines show the results excluding background noise. The SPDC source is adjusted to keep a constant \mbox{$\mu_s(\Delta \lambda) = 0.005$}. (b) Signal and idler filter heralding efficiencies for flat-top (red) and Gaussian (blue) filters and spectral purity for flat-top (solid line) and Gaussian (dashed line) filters.}
\label{filtershape}
\end{figure}
Having explored the effects of filtering for our particular experiment, here we perform simulations to investigate other considerations in the design of sources and filters that can apply more generally.\\
\indent We simulate the impact of the pump pulse width for photon pair generation, which will change the unfiltered JSA and the behavior of the FHE versus filter bandwidth. Flat-top spectral filters were also shown to provide a higher FHE compared to Gaussian shaped filters \cite{silberhorn}, which indicates that the filter shape could impact performance in the context of this study. Furthermore, we investigate the role of multiphoton pair emission in each case to gain more insight to its impact in both noisy and noiseless environments. We estimate the single-mode purity for each case to compare the noise impact on two-fold coincidence fidelity with the raw HOM visibility that could be achieved with the filtered photons.\\
\indent In Fig.~\ref{filtershape}, we compare the effects of changing the bandwidth of Gaussian or flat-top filters in noisy and noise-free environments. We simulate the CAR of a photon-pair source transmitting 1536.5-nm photons over the same 25-km/25-km fiber spools with the 1547.72-nm classical signals set to $P_0 = -12.5$\,dBm. Temporal filtering is fixed at $\Delta T =300$\,ps. In this simulation, we adjust the SPDC pump power so that $\mu_s(\Delta \lambda) = 0.005$ for each bandwidth setting such that the SNR improves as the filter is restricted, which is in contrast to our experimental measurement in Fig.~\ref{CARbw}. 
\begin{figure*}[!t]
\centering
%\onecolumn
\includegraphics[width=\textwidth]
{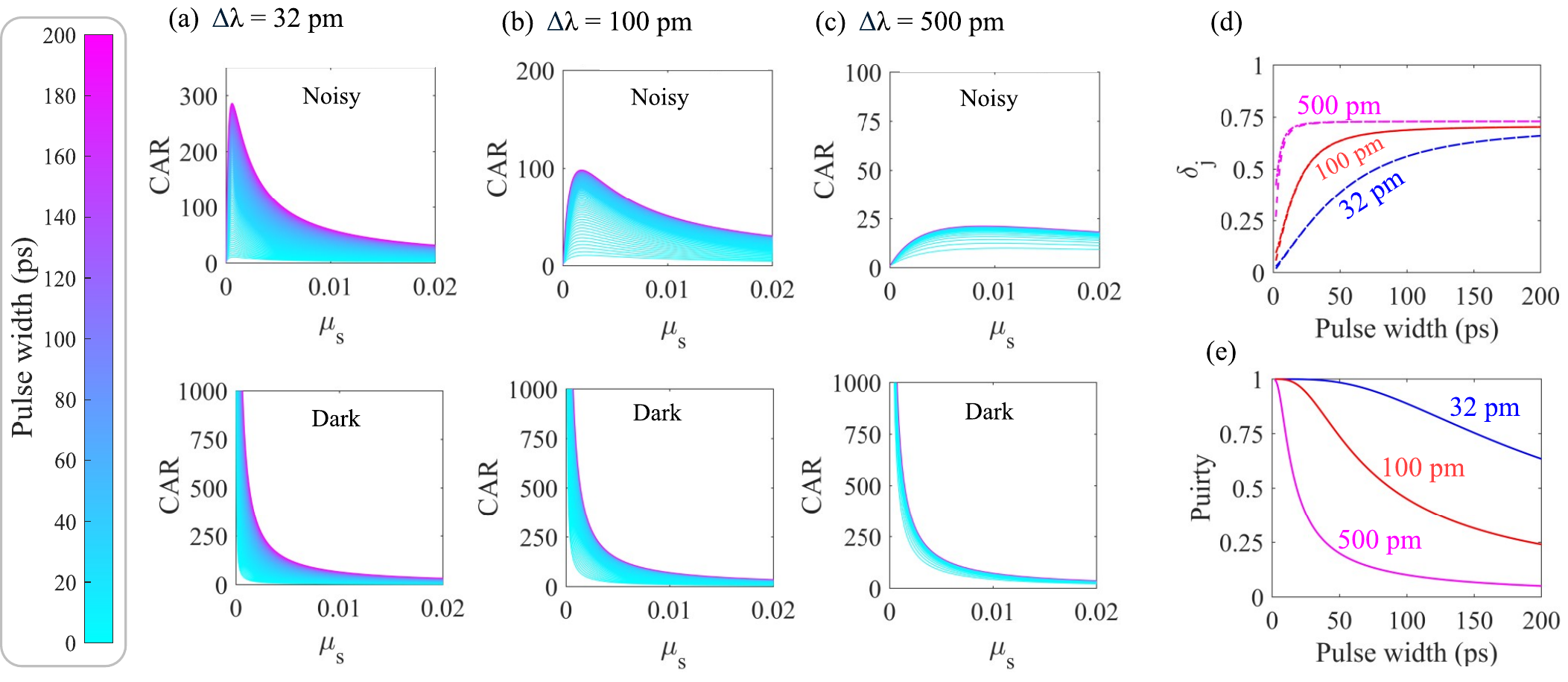}
\caption{Simulation of the impact of a source's pump pulse width ($\tau_p$) relative to $\Delta \lambda$ as a function of $\mu_s$. We simulate both noisy (top) and background noise-free (bottom) scenarios, where the noisy case has $P_0 =-12.5$\,dBm in each 25-km fiber. (a-c) CAR versus $\mu_s$ for (a) $\Delta \lambda_{s,i}=$\,32\,pm, (b) 100\,pm, or (c) 500\,pm, where $\Delta T_{s,i}=300$\,ps. Each simulation is color-coded according to a pump pulse width (left), ranging from $\tau_p=2$\,ps (teal) to 200\,ps (purple). The CAR significantly depends on $\tau_p$ and $\Delta \lambda$, which is due to how the FHE ($\delta_j(\tau_p, \ \Delta \lambda_{j})$) changes based on the relationship between these parameters (d). (e) Estimated spectral purity versus $\tau_p$ for each $\Delta \lambda$ to relate to the compatibility of each with HOM interference-based systems.}
\label{CARsim}
\end{figure*}
However, coincidence experiments have a more nuanced behavior. Fig.~\ref{filtershape}(a) shows the CAR as we vary $\Delta \lambda$ for both noisy (solid lines) and noiseless (dashed lines) scenarios. The black lines are the results for an ideal source that has a constant $\delta_j (\Delta \lambda) = 1$. Fig.~\ref{filtershape}(b) shows a simulation of $\delta_j (\Delta \lambda)$ for each filter type and the estimated spectral purity.\\
\indent Fig.~\ref{filtershape}(a) demonstrates that flat-top filters achieve a higher CAR when background noise is or is not present when $\delta_{\rm PS} <1$. In the noisy case, we see that the filter bandwidth needs to be <1\,nm to filter out enough noise for the CAR to increase to practical values; however, reducing the bandwidth more than this begins to decrease the FHE (Fig.~\ref{filtershape}(b)). Interestingly, we see that the CAR begins to decrease around $\Delta \lambda < 0.05$\,nm, even though the single-detector SNR continues to increase with the rejection of more SpRS noise. This is attributed to the increased sensitivity to multipair emission when the FHE is imperfect, which can be seen by the simulation that excludes background noise in Fig.~\ref{filtershape}(a) (dashed lines), which decreases according to the FHE in Fig.~\ref{filtershape}(b). To correct for this, one would need to reduce the SPDC pump power, which means that even in ideal environments, $\delta_{\rm PS}<1$ affects the coincidence rate \textit{and} the CAR. In a noisy environment, attenuating the pump would mean reducing the SNR at each detector, showing a more nuanced trade-off between maximizing the SNR and minimizing degradation due to multipair emission (see Appendix~\ref{appendix_model}). The flat-top filters outperform the Gaussian filters mainly due to the higher FHE versus noise rejection. Importantly, flat-top filters achieve the highest CAR for a given spectral purity, which implies that they are more optimal for $N$>2-photon applications.\\
\indent In Fig.~\ref{CARsim} we investigate the CAR as a function of the source's pump pulse width ($\tau_p$), which changes how the FHE varies with filter bandwidth ($\delta_j(\tau_p, \ \Delta \lambda_{j})$), and the pump power, which varies the impact of multiphoton pair emission. This is similar to our measurements in Fig.~\ref{300_50}(c), where the FHE was varied without changing the background noise, and has implications for the design of high-rate pulsed sources that require short pump pulses and filtered photons. Fig.~\ref{CARsim}(a-c) simulates the CAR for a source pumped by transform-limited Gaussian pump pulses between $\tau_p=2$\,ps to 200\,ps (FWHM) for both the noisy case (top) and noise-free case (bottom) for $\Delta \lambda_j = 32$\,pm (a), 100\,pm (b), or 500\,pm (c). We calculate the relevant parameters for each generated JSA using a Gaussian approximation developed in refs.~\cite{JSI_model, silberhorn} for the pump, phase matching, and filter functions. We neglect any effects of dispersion to maintain a more general analysis.\\
\indent Fig.~\ref{CARsim} clearly shows that performance is highly dependent on how the source is pumped both in intensity and in pump pulse width relative to $\Delta \lambda$. This is because $\delta_j(\tau_p, \ \Delta \lambda_{j})$ decreases more rapidly as a function of $\Delta \lambda_{j}$ for short pulses compared to longer pulses (Fig.~\ref{CARsim}(d)), which is due to the broadening of the pump spectral bandwidth and thus the width of the unfiltered JSA. This means that as the pump is shortened, the FHE becomes problematic at wider filter bandwidths, wherein the SpRS noise photons are not as tightly filtered.\\
\indent Although this implies for this case study that optimal performance occurs when one increases $\tau_{p}$ while minimizing $\Delta \lambda$, this will notably reduce the spectral purity of the photons (Fig.~\ref{CARsim}(d)), in which case optimization of parameters depends on the application. This again shows a trade-off between the two-fold fidelity and the optimality for applications such as state teleportation and entanglement swapping. {We note that narrow pump pulses could allow tighter temporal filtering without cutting into the joint temporal correlations, which could change this comparison if we simulated a variable $\Delta T$ according to $\tau_p$ and the filter coherence time. However, for the narrower pulses, this is conditioned on having ultrahigh-resolution detectors.}\\
\indent In the noise-free simulations at the bottom of Fig.~\ref{CARsim}(a-c), we see that the CAR as a function of the source intensity also depends on the relationship between $\tau_p$ and $\Delta \lambda$, with short pulses having a lower CAR for the same number of single photons. This again indicates the importance of mitigating accidental coincidences between higher-order Fock states occupying incompatible frequencies (see Appendix~\ref{appendix_model}).\\
\indent An important conclusion from these results is that an imperfect FHE will cause a system to be more susceptible to noise that is independent of the filter bandwidth, such as intrinsic detector dark counts or pump leakage. For each plot in Fig.~\ref{CARsim}(a-c), the background noise rate was kept constant, but the fidelity still varied according to the FHE. For example, this implies that an imperfect FHE limits the maximum distance or attenuation a quantum signal can experience when limited by intrinsic detector noise.\\
\indent To get a quantitative sense of exactly how much the CAR is affected due to the FHE, we can derive a simple equation for the maximum CAR that can be achieved after optimizing the source's pump power to balance multipair emission and SNR for a given $\delta_{j}$, $\eta_j$, and $\mathcal{D}_j$, where $\mathcal{D}_j$ is a noise count probability representing an arbitrary background noise source. We find that maximum fidelity and $\mu_{\mathrm{both}}^{\mathrm{opt}}$ are both directly related to the PSHE, $\delta_{\rm PS} \equiv \sqrt{\delta_s \delta_i}$. Maximizing with respect to $\mu_{\rm both}$ yields the following expression for the maximum CAR (see Appendix~\ref{appendix_model}):
\begin{equation}
\mathrm{CAR}_{\max }=\frac{\delta_{\mathrm{PS}} \sqrt{\eta_s \eta_i \mathcal{D}_s \mathcal{D}_i}}{2 \mathcal{D}_s \mathcal{D}_i+\delta_{\mathrm{PS}} \sqrt{\frac{\mathcal{D}_s \mathcal{D}_i}{\eta_s \eta_i}}\left(\frac{\eta_s \mathcal{D}_i}{\delta_i}+\frac{\eta_i \mathcal{D}_s}{\delta_s}\right)}+1,
\end{equation}
\noindent which is maximized when $\mu_{\mathrm{both}}^{\mathrm{opt}}= \delta_{\mathrm{PS}}\sqrt{{\mathcal{D}_s \mathcal{D}_i }/{\eta_s \eta_i}}.$ As an example, in the special case of $\delta_s \approx \delta_i \equiv \delta_{\rm PS}$, then both $\mathrm{CAR}_{\max }$ and $\mu_{\mathrm{both}}^{\mathrm{opt}}$ simply decrease linearly with $\delta_{\rm PS}$. If the FHE of each photon is measured as $\delta_{\rm PS} \approx 0.2$, for example, then the maximum CAR and the coincidence rate that it occurs are 20\% of that which would be obtained with an ideal PSHE. For an entangled source with a maximum TPI visibility of 96\% when $\delta_{\rm PS} = 1$, $\delta_{\rm PS} = 0.2$ would limit visibility to only 81.5\%. This clearly shows that careful attention to the design of sources and filters can have a significant impact on the achievable rates and fidelity in noisy environments.
\section{Discussion and Outlook}\label{conlusion_and_outlook}
In summary, we have studied various methods of filtering photon pairs generated from SPDC/SFWM sources in both noisy and ideal conditions, which has broad implications for optimal source and receiver designs for many multiphoton quantum applications. We further conducted the first study of pulsed time-bin entanglement distribution coexisting with classical communication in fiber, discussing some unique considerations for optimal sources, filters, and qubit measurement systems. Using narrow filtering, we demonstrated the co-propagation of C-band time-bin entangled photons and 10-Gbps C-band classical data over a total of 50\,km of standard fiber and showed that our system could tolerate mW-level C-band power. Although immediately applicable to entanglement-based applications, the study of two-fold fidelity alongside single-mode purity provides valuable insight on the consequences of filtering for more complex teleportation-based applications.\\
\indent Regarding the design of optimal sources for $N$>2-photon applications, it is important to maximize high single-mode purity, PSHE, and noise rejection simultaneously. Based on the results in this paper, the optimal unfiltered JSA produced at the source for spectral filtering is factorable such that the generated photons are pure without filtering, but \textit{both} photons are narrowband such that they can pass noise filters without rejecting source photons. Most studies attempting to resolve the trade-off between FHE and purity focus on engineering sources to avoid external filtering \cite{FHE_elim_freq_space_Grice_Uren_Walmsley_2001, FHE_garay2007photon2007, FHE_ultrafast2008, FHE_halder2009nonclassical, FHE_theorycavity_2010, FHE_beamsilb_2011, FHE_dualpumpFWM_2013, FHE_cavity_Rielander_etal_2016, FHE_Paesani2020NearIdealPhotonSources, Liu2020_highSpectralPurityPhoton, multistage2020, FHE_Xin_dispEng_LithNio_2022, psiquantum}, but not all JSAs may satisfy the narrow filtering requirement. For example, one approach is to engineer a nonlinear medium such that the group velocity of the photons match the pump. However, in some cases one photon can have a broader bandwidth than the other, potentially limiting the filtering around both without impacting the PSHE. Another approach is to use cavity-enhanced sources to design ultra-narrow bandwidth sources \cite{FHE_theorycavity_2010, FHE_cavity_Rielander_etal_2016, ErQMtele2025}, in which case noise can be narrowly filtered around one of the cavity modes. Significant progress has also been made with appropriately designed integrated photonic sources \cite{FHE_Paesani2020NearIdealPhotonSources, Liu2020_highSpectralPurityPhoton, psiquantum}, which can, in principle, also have narrow bandwidths and PSHE. Beyond SPDC/SFWM, atom-like entangled-photon sources could instead be used due to the ultra-narrow spectrum, which has been demonstrated using quantum dots in noisy free-space experiments \cite{BassoBasset_et_al_2023}.\\
\indent Instead of engineering the source, coherent filtering could help resolve these issues while offering greater noise tolerance. Coherent filtering includes mode-selective detection \cite{QPG_2011, yupingFilt2017, RayFilter2020}, such as quantum frequency conversion with an appropriately tailored pump \cite{QPG_2011, yupingFilt2017}. By passing only a single spectrotemporal mode per detector, the PSHE and noise rejection could be simultaneously maximized assuming ideal mode-selection efficiency. However, incoherent filtering has been the most common approach in quantum-classical network integration and free-space experiments to date because of the simple implementation and lower cost. We note that similar concepts apply to the measurement of continuous variable quantum states, which has been shown to significantly mitigate the impact of SpRS noise from classical communications \cite{CV_CC_2015}.\\
\indent For time-domain filtering, we note that ultra-narrow spectral bandwidths place limits on the repetition rate because of the temporal length of the photons. Even more relevant, the spectral bandwidth also governs how narrow a time-domain filter can be before cutting into the joint temporal amplitude or intensity. Instead of or in combination with short pump pulses, independent photons can be interfered by detecting photons within a narrower time resolution than the filter's coherence time \cite{GHZ_first_1997, Halder_etal_2007}. Using Fourier transformations, our model could be generalized to account for narrow frequency-time filtering.\\
\indent Beyond this, the model could be expanded to more accurately predict the role of multipair emission in the high-gain regime, a multimode analysis, and experiments involving more than two photons. Our results imply that for multiple pair sources that each have the same PSHE, $N$-fold coincidences would approximately scale as $C_N \propto \delta_{\rm PS}^{N/2}$ with visibility negatively affected by multipair emission according to $\mu_{s/i}>\mu_{\rm both}$ for each source.\\
\indent From the perspective of advancing the field of coexisting quantum and classical fiber networks, Tbps-level C-band classical data rates could have been communicated over each 25-km fiber by spreading the power across multiple attenuated DWDM channels and replacing our transceivers with more power-efficient devices \cite{Tbps_CC_narrowfilt_2024}. In fact, SpRS could be more suppressed using different C-band classical wavelengths, since our 1536.5-nm/1547.72-nm channel selection is within a Raman gain peak in the C-band SpRS spectrum \cite{eraerds_quantum_2010_2, Tbps_CC_narrowfilt_2024}.\\
\indent Importantly, our results apply to any wavelength scheme for WDM quantum-classical integration, such as systems using O-band quantum or classical sources. For example, using wavelength engineering to obtain weaker SpRS could enable wider band filtering, which could improve the FHE or performance of high-rate pulsed sources. The experimental design also provides valuable information on how networks of erbium-doped quantum memories could be impacted by classical signals due to the 1536.5-nm wavelength of both entangled photons \cite{Er2010, ErQMtele2025}. Based on our results, it is promising that these quantum networks could feasibly operate alongside high-rate classical communications, but further work is needed to evaluate these challenges due to the strong SpRS observed here and the unique physics.\\
\indent Although we used quantum-classical networking as an experimental case study, we believe that our results will be of general interest for other noisy environments such as ambient light in free-space quantum applications \cite{freespace1, Freespace_2007_2, FreespaceFilt_2018, bouchard_achieving_2021_2, BassoBasset_et_al_2023, freespace_2024} or noise due to SpRS from the pump in SFWM-based photon pair sources \cite{kumarSFWM, noiseSFWM}, quantum frequency conversion devices \cite{QFCnoise:18,  braggQFC2024}, and all-optical switches or shutters \cite{premNOLM, cameronXPM, Kupchak:19}. Furthermore, the analysis of multipair emission when the PSHE is imperfect could be useful for general experiments regardless of additional background noise.

\begin{acknowledgments}
This work is supported by the Fermi Forward Discovery Group LLC under Contract No. FWP-23-24 with the U.S. Department of Energy, Office of Science, Advanced Scientific Computing Research (ASCR) Program.
\end{acknowledgments}

\bibliography{ref}% Produces the bibliography via BibTeX.

\appendix

\section{Characterization of frequency and time-domain filters}\label{appendix_section1}

\begin{figure}[!b]
\centering
\includegraphics[width=1\linewidth]{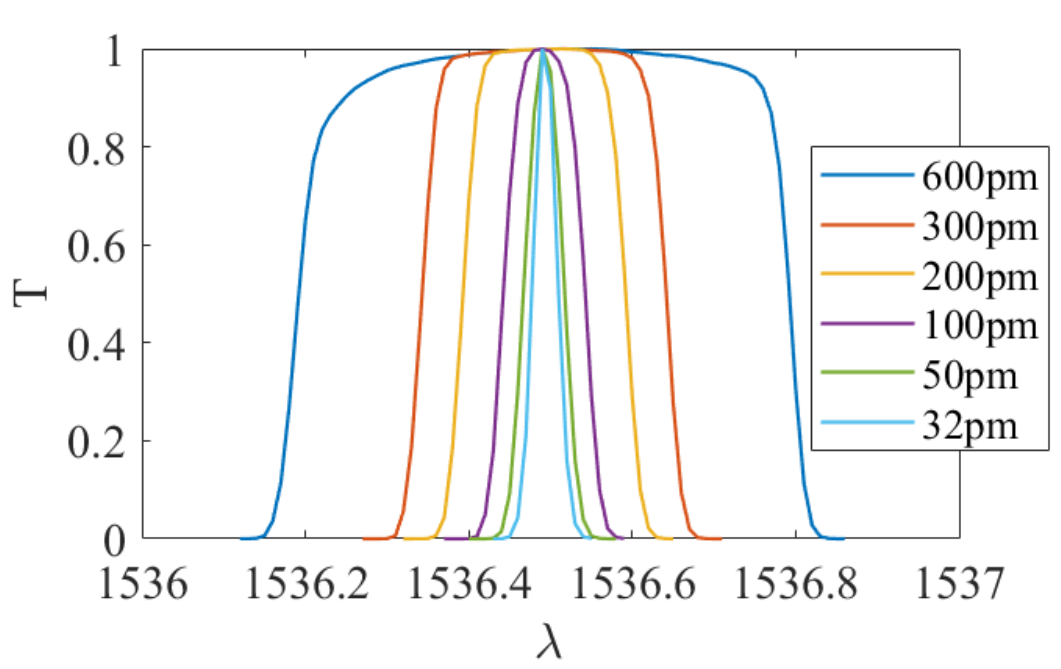}
\caption{Characterization of the shape of the tunable bandwidth filters as a function of bandwidth. }
\label{filters}
\end{figure}

Here we characterize the properties of our tunable filter passbands used in the experiment for spectral filtering. Fig.~\ref{filters} shows the normalized filter transmission of various bandwidth settings used in the experiment. We see that the filters are approximately flat-top with a slight slope at wider bandwidths. At narrow bandwidths they approach a more Gaussian-like shape due to the slope of the edge of the filter shape.
\begin{figure}[!t]
\centering
\includegraphics[width=1\linewidth]{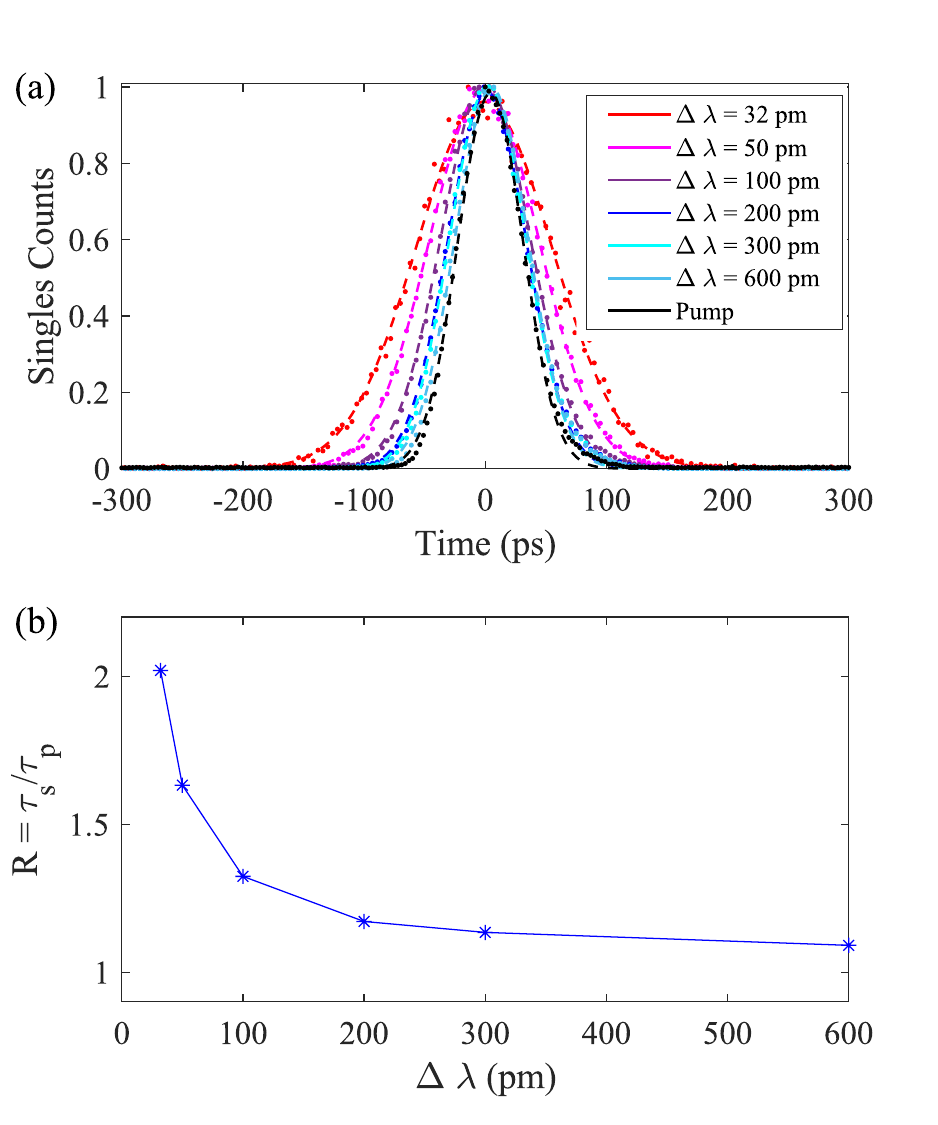}
\caption{(a) Normalized singles counts relative to the clock arrival time for different settings of $\Delta \lambda$. The black curve shows the 768.25-nm SPDC pump pulse when attenuated to the single-photon level and injected into the same SNSPD. (b) Ratio of the signal photon's temporal width relative to the pump pulse ($R = \tau_{s}/\tau_p$) versus $\Delta \lambda$.}
\label{cohtime}
\end{figure}
We also characterize the filtered photons in the time domain. We plot the broadening of the photon's coherence time relative to the source's pump width as a function of reducing $\Delta \lambda$. We first measure time-domain traces of the signal photon's temporal shape for filter settings ranging from $\Delta \lambda = 32$\,pm to 600\,pm (Fig.~\ref{cohtime}(a)). We then attenuate the pump pulse that drives SPDC to the single-photon level and directly send this pulse to the same SNSPD. It is clear from Fig.~\ref{cohtime}(a) that as the filter is restricted, the SPDC photons begin to broaden wider than the pump, and approach the pump width as the filter is widened. Fig.~\ref{cohtime}(b) shows the ratio between the FWHM of the filtered photons relative to the measured pump pulse. This is the characteristic property needed to interfere photons from independent SPDC/SFWM sources \cite{ind_photon}. Fig.~\ref{cohtime}(a) shows that for $\Delta \lambda =50$\,pm, which is what we used for the time-bin entanglement study, a 200-ps coincidence window is just narrow enough to tightly filter noise without cutting into the joint temporal correlations or source counts.
\section{Impact of multiphoton pair emission after spectral filtering}\label{appendix_model}
To model these experiments, it is critical to include the effects of higher-order Fock state emission during SPDC or SFWM. The spontaneous parametric processes of SPDC or SFWM both generate a two-mode squeezed vacuum (TMSV) state \cite{TMSV885}. In the Fock state basis, a single-mode TMSV source has the quantum state (before loss) of $|\xi\rangle_q=\sum_{m=0}^{\infty}(-1)^m \sqrt{\mu^m /\left(1+\mu\right)^{m+1}}\left|m_s, m_i\right\rangle$. Here, $m$ are the number of photons in signal and idler modes, which follows a thermal photon number distribution as a function of the mean number of photon pairs per pulse $\mu$ produced in the medium: \mbox{$P_{th}(\mu, m) = {{\mu^m }/{\left(1+\mu\right)^{m+1}}}$} \cite{TMSV885}. For our purposes, only the $m=1$ term is desired, where $m>1$ leads to accidental coincidence counts or non-zero coincidences in an entangled photon pair's TPI fringe \cite{takesue_effects_2010_2, Takeoka_2015}. Many sources, especially when not filtered, have many modes emitting $\mu_k$ mean number of photons per mode. A multimode source with $M$ modes has a photon number distribution ranging from thermal ($M=1$) to Poisson ($M \rightarrow \infty$) as a function of $\mu = \sum \mu_k$, since it is a convolution of $M$ thermal sources. However, a Poisson model is a good approximation to predict the impact of multiphoton emission when $\mu_{} \ll 1$ for any number of modes \cite{takesue_effects_2010_2}, which significantly simplifies modeling. Since we operate at $\mu \ll 1$ in our experiment, we build on this approximation and leave a full multimode analysis in the high-gain regime to future work.\\
\indent When $\mu$, the efficiency in signal/idler channels $\eta_j$, and total background noise count probability $\mathcal{D}_j$ are all $\ll 1$, the single-detector ($S_j$), coincidence ($C$), and accidental coincidence ($A$) count probabilities per detection window for a SPDC/SFWM experiment are often modeled as \cite{Takesue2006_TimeBinDistribution,  takesue_effects_2010_2, Takeoka_2015}:
\begin{equation}\label{Beq1}
S_j=\mu \eta_j+\mathcal{D}_j,
\end{equation}
\begin{equation}\label{Beq2}
C = \mu \eta_s \eta_i + S_s S_i,
\end{equation}
\begin{equation}\label{Beq3}
A=S_s S_i.
\end{equation}
\indent However, these equations neglect the properties of the source's unfiltered JSA, $f(\omega_s, \omega_i)$, and the impact of external filtering. The approach in this paper is to note that imperfect FHE means that the number of photons that can lead to signal, idler, and coincidence detections are unequal.\\ 
\indent To include this, we build on refs.\,\cite{silberhorn, highrate_timebin} and propose the definitions for the mean number of photons that pass signal ($j=s$) or idler ($j=i$) filters as
\begin{equation}
\mu_j=\mu_T \iint d \omega_s d \omega_i\left|f\left(\omega_s, \omega_i\right) g_j\left(\omega_j\right)\right|^2,
\end{equation}
\noindent and the number of photons that pass both filters
\begin{equation}
\mu_{\mathrm{both}}=\mu_T \iint d \omega_s d \omega_i\left|f\left(\omega_s, \omega_i\right) g_s\left(\omega_s\right) g_i\left(\omega_i\right)\right|^2,
\end{equation}
\noindent where $\mu_T$ is the mean number of photons generated when the source is not filtered and $g_j\left(\omega_j\right)$ are functions representing the signal/idler normalized filter transmission functions. Using these definitions, we modify the commonly used equations to include the effects of filtering a source's joint spectrum.\\
\indent The two-fold coincidence probability per pulse is modeled by subsituting $\mu \rightarrow \mu_{\rm both}, \mu_s, \mu_i$ accordingly into Eq.~\ref{Beq2} to obtain
\begin{equation}
C \approx \mu_{\mathrm{both}} \eta_s \eta_i+S_s S_i,
\end{equation}
\noindent where the single-detector count probability is obtained by substituting $\mu \rightarrow \mu_s$ and $\mu \rightarrow \mu_i$ in Eq.~\ref{Beq1} to obtain:
\begin{equation}
S_s=\mu_s \eta_s + \mathcal{D}_s
\end{equation}
\begin{equation}
S_i=\mu_i \eta_i + \mathcal{D}_i.
\end{equation}
\indent Here $D_j$ is the total background noise count probability, which we modeled in more detail in the main text to account for the physics of our specific experiment.\\
\indent Accidental coincidences are modeled similar to Eq.~\ref{Beq3} but using the new definitions for $S_s$ and $S_i$, which is given by the product of signal and idler single-photon probabilities
\begin{equation}
A=S_s S_i.
\end{equation}
\indent When $\mu_{j} = \mu_{\rm both}$, the FHE is ideal, which gives the typical equations for the probability of coincidence and singles counts \mbox{$C=\mu \eta_s \eta_i + S_s S_i$} and \mbox{$S_j = \mu \eta_j + \mathcal{D}_j$} in refs.~\cite{Takesue2006_TimeBinDistribution,  takesue_effects_2010_2, Takeoka_2015}.\\
\indent In a noisy environment, the CAR is given by
\begin{equation}\label{CARnoisyfull}
\mathrm{CAR}=\frac{\mu_{\rm both} \eta_s \eta_i}{\left(\mu_{s } \eta_s+\mathcal{D}_s\right)\left(\mu_{i} \eta_i+\mathcal{D}_i\right)}+1.
\end{equation}
\indent It can be useful to arrange this using only one parameter to account for the source intensity. Since \mbox{$\mu_{i/s} = \mu_{\rm both}/\delta_{s/i}$}, we substitute this into eq.~\ref{CARnoisyfull} to obtain:
\begin{equation}\label{muBCAR}
\mathrm{CAR}=\frac{ \mu_{{\rm both}} \eta_{{s}} \eta_{{i}}}{\left(\delta_i^{-1}{\mu_{{\rm both}}}{} \eta_{{s}}+\mathcal{D}_{{s}}\right)\left(\delta_s^{-1}{\mu_{{\rm both}}}{} \eta_{{i}}+\mathcal{D}_{{i}}\right)}+1.
\end{equation}
\indent Taking the derivative of this function with respect to $\mu_{\rm both}$, we can find the mean photon number that maximizes the CAR by balancing improvements to the SNR when increasing $\mu_T$ with the inevitable degradation of CAR caused by multipair emission.\\
\indent Ref.~\cite{silberhorn} proposed a parameter called the "pair-symmetric filter heralding efficiency" (PSHE), \mbox{$\delta_{\mathrm{PS}} \equiv \sqrt{\delta_s \delta_i}$}, which quantifies the impact of the FHE of both photons. Using this definition, the optimal $\mu_{\mathrm{both}}$ which maximizes the CAR can be expressed as
\begin{equation}
\mu_{\mathrm{both}}^{\rm opt}= \delta_{\mathrm{PS}}\sqrt{\frac{\mathcal{D}_s \mathcal{D}_i }{\eta_s \eta_i}}.
\end{equation}
\indent Substituting this into eq.~\ref{muBCAR}, this gives the maximum achievable CAR given a particular channel efficiency, background rate, and FHE of:
\begin{equation}
\mathrm{CAR}_{\max }=\frac{\delta_{\mathrm{PS}}\sqrt{\eta_s \eta_i \mathcal{D}_s \mathcal{D}_i}}{2 \mathcal{D}_s \mathcal{D}_i+\delta_{\mathrm{PS}}\sqrt{\frac{\mathcal{D}_s \mathcal{D}_i }{\eta_s \eta_i}}\left(\frac{\eta_s \mathcal{D}_i}{\delta_i}+\frac{\eta_i \mathcal{D}_s}{\delta_s}\right)} +1.
\end{equation}
\indent Thus, the maximum CAR that can be achieved clearly depends on $\delta_{\rm PS}$, and the coincidence rate in which the CAR is maximized also decreases linearly with the PSHE.\\
\indent A common method for using experimental data to obtain or predict key parameters for photon pair experiments is to use the experimental result for the CAR in noiseless environments, $\rm CAR_{\rm dark}$, to obtain either the mean photon number generated in the source. This is often done using equation $\rm CAR_{\rm dark} \approx 1 + 1/\mu$, which when rearranged gives $\mu \approx 1/({\rm CAR_{\rm dark}-1})$ \cite{takesue_effects_2010_2}. One can also predict the CAR as a function of $\mu$ using this simple relationship.

However, when $\delta_{\rm PS}<1$, this is no longer valid. In the case of imperfect FHE in the low-gain regime, we find that the CAR in noise-free environments is related to the pair generation probability by the equation
 \begin{equation}\label{xx1}
      \rm CAR_{\rm dark}  = \frac{\mu_{\rm both}}{\mu_{s} \mu_{i}} +1.
\end{equation}
\indent Using the relationship $\mu_{i/s} = \mu_{\rm both}/\delta_{s/i}$, we can write this equation as
 \begin{equation}\label{CARPSH}
      \rm CAR_{\rm dark}  = \frac{\delta_s \delta_i}{\mu_{\rm both}} +1,
\end{equation}
\noindent which can be used to estimate $\mu_{\rm both}$ using experimentally measured parameters for the CAR and FHEs. Eq.~\ref{CARPSH} can be useful to predict the CAR based on the pair generation probability. This shows that for a given coincidence rate ($\propto \mu_{\rm both}$) the CAR is reduced by a factor of $\delta_{\rm PS}^2$ compared to the ideal case.
\subsection{Equal heralding efficiency approximation}
\begin{figure*}[!t]
\centering
%\onecolumn
\includegraphics[width=\textwidth]
{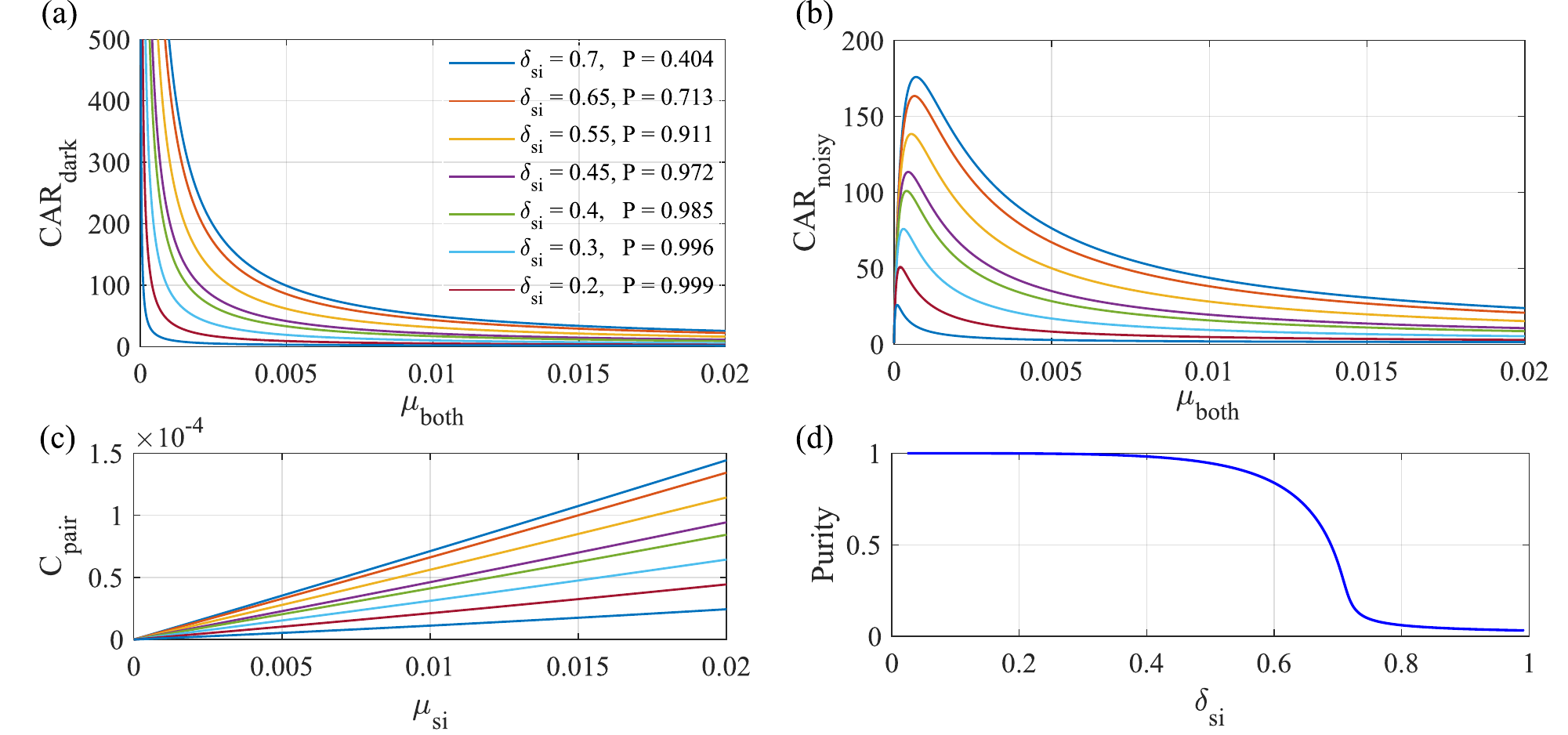}
\caption{(a) CAR versus $\mu_{\rm both}$ in a noise-free environment for different values of $\delta_{\rm si} \equiv \delta_s = \delta_i$. (b) CAR when the signal/idler background count probability is $\mathcal{D}_j = 10^{-4}$ and $\eta_j = 0.1$. (c) Single-pair coincidence probability versus the number of generated signal/idler photons, $C_{\rm pair} = \delta_{\rm si} \mu_{si}\eta_s \eta_i$, for different values of $\delta_{\rm si}$. (d) Simulated single-mode purity as a function of $\delta_{\rm si}$.}
\label{CAR_delta}
\end{figure*}

If we assume that the FHE for each photon is approximately equal ($ \delta_s = \delta_i  \equiv \delta_{\rm PS}$), which is not universally true and depends on the source's phase matching angle and bandwidth \cite{silberhorn} or the similarity between signal and idler filters, this implies that $\mu_s \approx \mu_i \equiv \mu_{si} $. This leads to the relationship $\mu_{\rm both} \approx \delta_{\rm PS}  \mu_{si}$. Substituting this into the model, we obtain the following equations:
\begin{equation}\label{approx1}
C \approx  \delta_{\rm PS}  \mu_{si} \eta_s \eta_i  + S_s S_i,
\end{equation}
\begin{equation}\label{approx2}
S_j \approx  \mu_{si} \eta_j   + D_j.
\end{equation}
\indent In this approximation, the CAR excluding any background noise simplifies to:
 \begin{equation}
\begin{gathered}
\mathrm{CAR}_{\text {dark }} \approx \frac{\delta_{\mathrm{PS}}}{\mu_{\mathrm{si}}}+1 \\
\approx \frac{\delta_{\rm PS}^2}{\mu_{\text {both }}}+1,
\end{gathered}
\end{equation}
\noindent which means that one can use the measured CAR to estimate the experimental parameters:
\begin{equation}\label{musi}
\mu_{s i} \approx \frac{\delta_{\rm PS}}{\mathrm{CAR}_{\text {dark }}-1}
\end{equation}
\begin{equation}
\mu_{\text {both }} \approx \frac{\delta_{\rm PS}^2}{\mathrm{CAR}_{\text {dark }}-1}.
\end{equation}
\indent Eq.~\ref{musi} resembles the one derived in ref.~\cite{highrate_timebin}, however our work expands this approach to more accurately quantify the number of photons which lead to singles and coincidence counts.\\
\indent Using the equal FHE approximation, the equation for the CAR accounting for background noise is then given by
\begin{equation}
\mathrm{CAR}=\frac{\delta_{{\rm PS}} \mu_{{si}} \eta_{{s}} \eta_{{i}}}{\left(\mu_{{si}} \eta_{{s}}+\mathcal{D}_{{s}}\right)\left(\mu_{{si}} \eta_{{i}}+\mathcal{D}_{{i}}\right)}+1,
\end{equation}
\noindent and the value for $\mu_{si}$ that maximizes this function is given by
\begin{equation}
\mu_{si}^{\rm opt} = \sqrt{\frac{\mathcal{D}_s \mathcal{D}_i}{{\eta_s \eta_i}}}.
\end{equation}
\indent Interestingly, $\mu_{si}^{\rm opt}$ does not depend on $\delta_{\rm PS}$, which means that the optimal number of signal/idler photons is roughly the same regardless of the value of the FHE.\\
\indent However, the maximum CAR that can be achieved for a given $\eta_j$, $\mathcal{D}_j$, and $\delta_{\rm PS}$ is given by
\begin{equation}
    \text{CAR}_{\rm max}(\delta_{\rm PS})=\frac{\delta_{\rm PS} \sqrt{\eta_s \eta_i \mathcal{D}_s \mathcal{D}_i}}{2 \mathcal{D}_s \mathcal{D}_i+\sqrt{\frac{\mathcal{D}_s \mathcal{D}_i}{\eta_s \eta_i}}\left(\mathcal{D}_s \eta_i+\mathcal{D}_i \eta_s\right)} +1.
\end{equation}
\indent The optimal $\mu_{\rm both}$ to maximize the CAR is still given by $\mu_{\rm both}^{\rm opt} = \delta_{\rm PS} \sqrt{\frac{{\mathcal{D}_s \mathcal{D}_i}}{{\eta_s \eta_i}}}.$ Thus, the single count rate that optimizes the CAR is independent of the PSHE, but the maximum CAR and the single-pair coincidence rate in which it is obtained are reduced by a factor of $\delta_{\rm PS}$.\\
\indent In Fig.~\ref{CAR_delta}, we use the approximation $\delta_{\rm PS} = \delta_s = \delta_i$ and plot the CAR as a function of $\mu_{\rm both}$ for different values of $\delta_{si}$ for dark (a) and noisy (b) environments. We analyze an example case with true channel efficiencies $\eta_s = \eta_i = 0.1$ and arbitrary background noise count probability of $\mathcal{D}_s = \mathcal{D}_s = 10^{-4}$. We also plot in (c) the single-pair coincidence probability as a function of the signal/idler photons $C_{\rm pair} = \delta_{\rm PS} \mu_{si}\eta_s \eta_i$. In (d), we plot the purity, ($P$) of the photons as a function of $\delta_{\rm PS}$ using a Gaussian approximation for the filters, pump, and phase matching function \cite{silberhorn} (see Appendix~\ref{app_gaussian_model}).\\
\indent As $\delta_{\rm PS}$ decreases, the source becomes generally much more susceptible to accidental coincidences caused by multiphoton emission at incompatible frequencies for single-pair coincidence detection (Fig.~\ref{CAR_delta}(a)). Compared to the ideal $\delta_{\rm PS}=1$, the CAR is reduced approximately a factor of $\delta_{\rm PS}$ for the same single-photon rate, and by $\delta_{\rm PS}^2$ for the same coincidence rate. Thus, not only does $\delta_{\rm PS}<1$ reduce the coincidence rate by $\delta_{\rm PS}$ for the same number of detected single photons, but also the CAR. To achieve the same CAR as $\delta_{\rm PS}=1$, one needs to reduce the pump power of the source until the true single-pair coincidence is decreased by a factor of $\delta_{\rm PS}^2$ compared to the ideal case, resulting in an even greater hit to coincidence rates when factoring in a desired target fidelity.\\
\indent Fig.~\ref{CAR_delta}(b) highlights the consequences of this effect in noisy environments. As $\delta_{\rm PS}$ decreases, the CAR decreases for all values of $\mu_{\rm both}$, even though noise and loss are unchanged. We can also see that as $\delta_{\rm PS}$ is reduced, $\mu_{\rm both}$ must be further attenuated to obtain its maximum CAR value. This is similar to the experiment in Fig.~\ref{300_50}(c) and the simulations in Fig.~\ref{CARsim}, since $\mathcal{D}_j$ is the same but $\delta_{\rm PS}$ is varied. Importantly, this is also similar to the impact of the intrinsic detector dark count noise, which does not change with either the pump or filter properties but its impact depends on $\delta_{\rm PS}$. This shows, for example, that reduced heralding efficiency can limit long-distance quantum communications excluding additional noise from coexisting classical signals.\\
\indent Fig.~\ref{CAR_delta}(d) shows the trade-off between fidelity due to filtering and single-mode spectral purity, which is directly related to the HOM visibility for independent sources of genuine single photons \cite{JSI_model}. Clearly, the impact of multipair emission as purity increases shows that maintaining high rates and high raw fidelity can be challenging, which is an important consideration when designing $N$-photon applications.
\section{Time-bin versus Polarization Entanglement Distribution in Noisy Environments}\label{appendix_timepol}
Here, we explore some differences between time-bin encoding and polarization encoding in noisy environments, including how time-bin qubits are measured. 
\subsection{Basis dependence in measuring time-bin qubits in high-noise environments}
\subsubsection{$X$ and $Y$ bases}
For time-bin entanglement encoding where the $X$ and $Y$ bases are measured using a 1-bit delay interferometer \cite{Takesue2008_TimeBinTomography}, the maximum and minimum coincidence counts based on the model in the main text for the middle bin are given by:
\begin{equation}
C_{\max }^{x y}=\frac{1}{4} \mu_{\mathrm{both}} \eta_s \eta_i\left(\frac{1}{2}+\frac{1}{2} V_{\mathrm{int}}\right) +S_s^{x y} S_i^{x y},
\end{equation}
\begin{equation}
C_{\min }^{x y}=S_s^{x y} S_i^{x y} + \frac{1}{4} \mu_{\mathrm{both}} \eta_s \eta_i\left(\frac{1}{2}-\frac{1}{2} V_{\mathrm{int}}\right),
\end{equation}
\noindent with the singles count probabilities of
\begin{equation}
S_j^{x y}=\frac{1}{2} \mu_j \eta_j+\left(\frac{1}{2} \eta_{r_j} \alpha_{\rm p o l} \mathcal{R}_j \Delta \lambda_j+d_j\right) \Delta T_j.
\end{equation}
\indent Compared to a correlated pair source, the coincidence rate is reduced by 1/4. This is because both signal and idler photons split and recombine at the first and second beam splitters of the interferometer, but 1/2 of the counts are lost to the two side bins. Due to the random arrival time for the SpRS photons, the noise is also reduced 50\% in each detector, since photons will randomly split and recombine at each splitter. This means that at the output, the noise is equal in all time intervals but with 50\% of the input intensity. For unpolarized noise, polarization filtering can reduce the noise count probability to 1/4 of the input since $\alpha_{\rm pol} = 0.5$.
\subsubsection{Optimal and non-optimal measurements in the $Z$ basis}
In our experiment, where the $Z$ basis is measured using the first and third bins of the interferometer output, the single counts are 1/4 of a correlated pair source, since these photons split at both beam splitters in the interferometer but do not recombine as for the middle bin. In this case, the coincidence probability is 1/16 of a correlated pair source. However, the noise photon count rate is equal across all three bins, indicating that the first and third bins are more impacted by noise than the $X/Y$ bases using this $Z$-basis measurement scheme.\\
\indent We model the maximum ($C_{t_1 t_1}$ or $C_{t_3 t_3}$) and minimum ($C_{\rm t_1 t_3}$ or $C_{\rm t_3t_1}$) coincidences for the $Z$ basis by the equations
\begin{equation}
    C_{\rm max} = \frac{1}{16}\mu_{\rm b o t h} \eta_s \eta_i+ S_s^z S_i^z,
\end{equation}
\begin{equation}
    C_{\rm min} = S_s^z S_i^z,
\end{equation}
\noindent and the singles count probability are given by
\begin{equation}
    S_{j}^z = \frac{1}{4} \mu_j \eta_j + \left(\frac{1}{2}\eta_{r_j}\alpha_{pol}\mathcal{R}_j \Delta \lambda_j+d_j\right) \Delta T_j.
\end{equation}
\indent Thus, it is clear that when one uses the first and third bins to measure the $Z$ basis, one will have an unequal degradation of fidelity when projecting onto qubits on the equator or poles of the Bloch sphere. This would, for example, lead to the reconstruction of a density matrix that resembles a non-maximally entangled state when performing quantum state tomography \cite{Takesue2008_TimeBinTomography}, even if the input Bell state is ideal when neglecting noise.\\
\indent There are some ways in which this can be addressed. For example, a beam splitter or switch could be placed in front of the interferometer (Fig.~\ref{compare}(a)), where some fraction of the photons are sent toward another detector without an interferometer to perform $Z$ basis measurements. In this case, the $Z$ basis would not suffer an additional source of loss compared to the noise and would maintain the same fidelity as the $X$ and $Y$ bases. Thus, it appears that using an interferometer for all three bases is not optimal for time-bin encoding in high-noise environments. For applications in which only two bases are required, the $X/Y$ bases would be optimal to achieve the highest fidelity or tolerance to background noise.
\subsection{Importance of polarization filtering and comparing to polarization encoding }

In this section, we compare time-bin entanglement to polarization entanglement and examine the benefit of polarization filtering of unpolarized noise photons.\\
\indent For polarization encoding, we build on the model in ref.~\cite{thomasx} but modify it to account for the properties of the source's JSA and filtering. For a source pumped in a coherent equal superposition of horizontal ($H$) and vertical ($V$) polarizations such that the number of photons generated across the JSI in each qubit degree of freedom is $\mu_T \equiv \mu_T^H = \mu_T^V$, the coincidence counts in the low-gain regime can be modeled by 
\begin{equation}
C_{\max }^{\rm pol}= \mu_{\mathrm{both}} \eta_s \eta_i +S_s^{\rm pol} S_i^{\rm pol},
\end{equation}
\begin{equation}
C_{\min }^{\rm pol}= S_s^{\rm pol} S_i^{\rm pol},
\end{equation}
\noindent where the singles count probability including background noise is
\begin{equation}
S_j^{\rm pol}=\mu_j \eta_j+\left(\eta_{r_j} \alpha_{\rm p o l} \mathcal{R}_j \Delta \lambda_j+d_j\right) \Delta T_j.
\end{equation}
\indent This shows that the coincidence probability is equivalent to the correlated pair source applying polarization filtering, since, for example, $\mu_j^H$ photons will pass a polarizer at the measurement apparatus projecting onto $|H\rangle$ with 100\% probability. For unpolarized noise, the input noise intensity is reduced by $\alpha_{\rm p o l}=1/2$ similar to polarization filtering with time-bin qubits due to the polarizer in the qubit analyzer. Thus, the analyzer naturally applies polarization filtering in the measurement procedure and has a $4 \times$ higher coincidence probability compared to time-bin encoding in the $X/Y$ bases \cite{takesue_effects_2010_2} and $16\times$ higher than the $Z$ basis measurements using interferometer outputs. Note that the coincidence probability for all bases are the same assuming unpolarized background noise photons and the generation and alignment of a maximally entangled Bell state.\\
\begin{figure}[!t]
\centering
\includegraphics[width=1\linewidth]{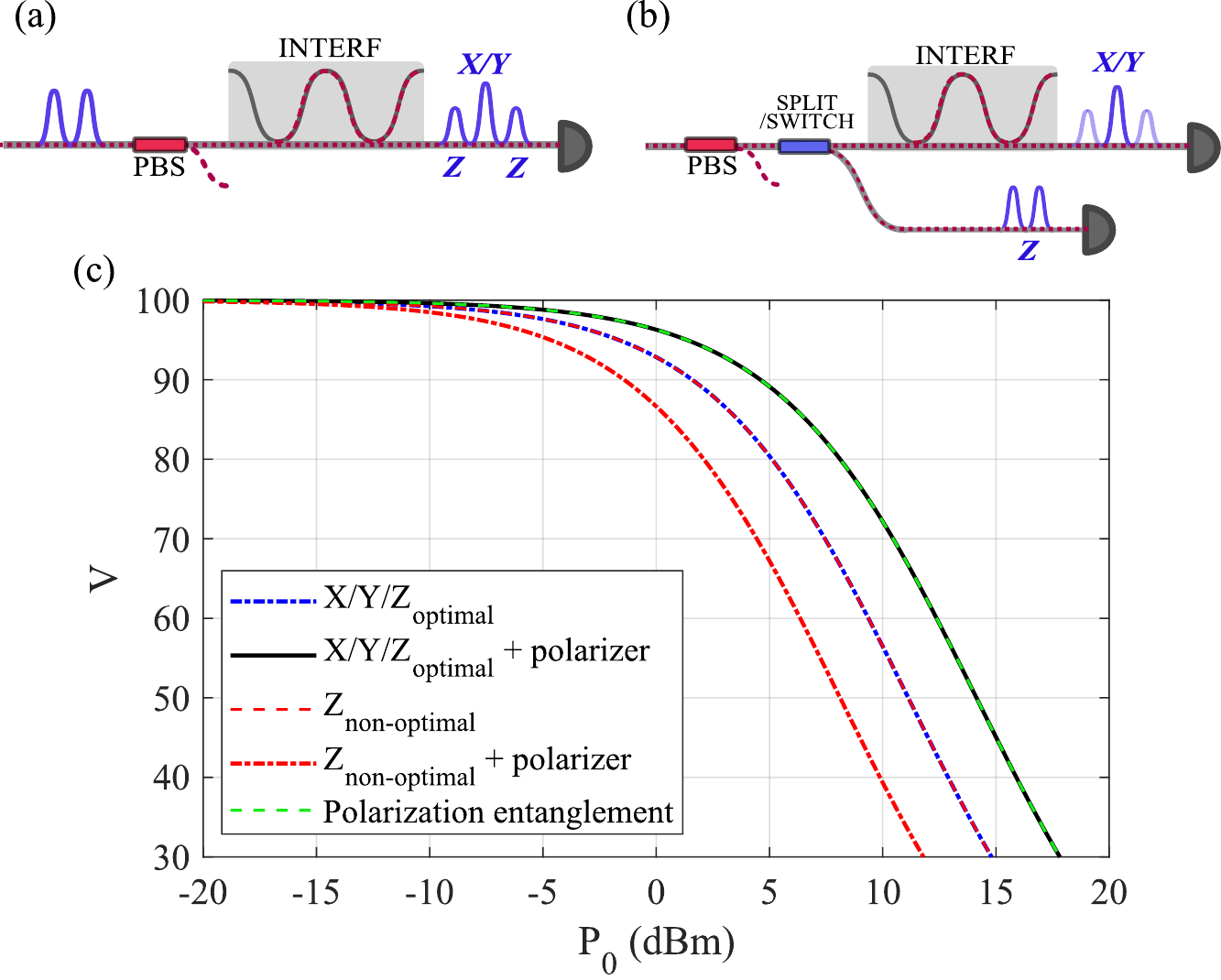}
\caption{Comparison between time-bin entanglement distribution and how the receiver is designed to polarization entanglement. (a) Time-bin encoding measured with a 1-bit delay interferometer, where all bases are measured using the interferometer output. (b) Time-bin encoding with a splitter or switch used to measure the $Z$ basis and interferometer for the $X/Y$ bases, which is the optimal case for the $Z$ basis. (c) Entanglement visibility as a function of classical power for each case assuming ideal sources and FHE.}
\label{compare}
\end{figure}
\indent To demonstrate the differences between these approaches and also evaluate the limits of source and filter designs, we simulate the performance of time-bin entangled photons compared to polarization entangled photons measured using a standard polarization analyzer. We further simulate an ideal source and receiver to see the limits of the tolerable classical power when all key parameters are optimized. We simulate both time-bin encoding using a 1-bit delay interferometer to measure all three bases (Fig.~\ref{compare}(a)) and time-bin encoding when one uses a splitter or switch to more optimally measure the $Z$ basis (Fig.~\ref{compare}(b)). For each case, we compare the performance with and without polarization filtering.\\
\indent In this simulation, we keep the same filtering used in our time-bin experiment (6.5\,GHz, 200\,ps) but assume an ideal FHE to examine the limits in an idealized case. We further assume no loss at the source node, which maximizes the SNR at each detector since this loss does not affect the noise photons generated in the long-distance fibers. The mean photon number at the source is optimized to find the highest possible visibility at each classical power according to the background noise level and channel losses. All other parameters are identical to those used in the 1536.5-nm entanglement distribution experiment over 25-km/25-km fibers with the co-propagating 1547.72-nm classical signals, representing the most ideal performance that could be achieved in principle with these filter settings.\\
\indent The results are shown in Fig.~\ref{compare}(c). We find that polarization encoding outperforms the time-bin entanglement distribution without polarization filtering. However, adding polarizers before the 1-bit delay interferometer makes time-bin encoding in the $X/Y$ bases as resilient to incident noise rates compared to polarization entanglement. However, the coincidence rate will still be 25\% of a polarization entangled state. When the $Z$ basis is measured at the interferometer output (non-optimal), the fidelity is more impacted and is unable to tolerate as high of classical power, which is what we observed in our experiment. However, if one uses a switch or splitter to measure the $Z$ basis without the interferometer (optimal), it can obtain the same tolerance to incident background noise as the $X$ and $Y$ bases. When polarization filtering is applied, all cases achieve the same visibility for all classical power levels.\\
\indent We also see that the tolerable classical power is much higher compared to the experimental demonstration, which is due to the increase of the source intensity to overcome the higher background noise, the ideal FHE, and the improvement of the loss before multiplexing with the classical signal. Using these improvements, we predict that it is feasible to reach C-band classical launch powers of around 10\,dBm in each fiber while still allowing entanglement-based quantum applications at C-band wavelengths.
\section{Gaussian Approximation for Modeling a Filtered JSA}\label{app_gaussian_model}
Here we briefly summarize the Gaussian approximation for the filtered JSA used in simulations of the CAR as a function of pump pulse width and filter bandwidth in the main text. We use the models in refs.~\cite{JSI_model, silberhorn}, where the JSA of a source is represented by the equation \cite{silberhorn}
\begin{equation}
\begin{aligned}
f\left(\omega_s, \omega_i\right) 
= & N \exp \left(\frac{-\left(\omega_s-\omega_{s 0}+\omega_i-\omega_{i 0}\right)^2}{4 \sigma_p^2}\right) \\
& \times \operatorname{sinc}\left(\frac{\left(\left[\omega_s-\omega_{s 0}\right] \sin \theta+\left[\omega_i-\omega_{i 0}\right] \cos \theta\right)}{2 \sigma_{p m}}\right).
\end{aligned}
\end{equation}
\indent Here, $\sigma_p$ is the pump's frequency bandwidth, $\sigma_{pm}$ is the phase matching bandwidth, and $\theta=\arctan \left(\frac{k_p^{\prime}-k_s^{\prime}}{k_p^{\prime}-k_i^{\prime}}\right)$ is the phase matching angle with $k_x^{\prime}$ being the frequency derivative of $k$ for the signal or idler photons. Filters can then be applied, which are modeled to have central frequencies of $\omega_{j 0}$.\\
\indent The filtered JSA can be expressed as \mbox{$\Phi(\omega_s, \omega_i) =  f\left(\omega_s, \omega_i\right) {g}_s\left(\omega_s\right) {g}_i\left(\omega_i\right)$}, where ${g}_j\left(\omega_j\right)$ are the normalized filter transmission functions. Then, each parameter could be obtained using numerical integration. Alternatively, simple analytical equations can be obtained for each parameter using a Gaussian approximation for the pump, phase-matching, and filter functions \cite{JSI_model, silberhorn}. To do this, the sinc function in the JSA is approximated by $\operatorname{sinc}(\mathrm{x}) \approx \exp \left(-\alpha x^2\right)$ with $\alpha=0.193$ such that the two have the same FWHM. Gaussian filters for the signal and idler photons can then be applied with bandwidths of $\sigma_{j}=\sigma_{j_{\rm F W H M}} /(2 \sqrt{2 \ln 2})$.\\
\indent Using this approximation, the equation for $\Gamma_{\rm both}$ which relates to the coincidence probability ($\mu_{\rm both} = \mu_T \Gamma_{\rm both}$) can then be derived as \cite{silberhorn} 
\begin{equation}
\Gamma_{\text {both }}  \approx \sqrt{\frac{a_0 b_0-c^2}{a b-c^2}},
\end{equation}
\noindent where
\begin{equation}\label{eqA5}
\begin{aligned}
a & =\frac{\alpha^2 \sin ^2 \theta}{\sigma_{p m}^2}+\frac{1}{\sigma_p^2}+\frac{1}{\sigma_s^2} \\
b & =\frac{\alpha^2 \cos ^2 \theta}{\sigma_{p m}^2}+\frac{1}{\sigma_p^2}+\frac{1}{\sigma_i^2} \\
c & =\frac{\alpha^2 \cos \theta \sin \theta}{\sigma_{p m}^2}+\frac{1}{\sigma_p^2}.
\end{aligned}
\end{equation}
\indent Setting one of the signal or idler filter bandwidths to infinity gives the marginal probabilities for each individual photon, giving the equations \cite{silberhorn}
\begin{equation}
\begin{aligned}
\Gamma_i & \approx \sqrt{\frac{a_0 b_0-c^2}{a_0 b-c^2}}, \\
\Gamma_s & \approx \sqrt{\frac{a_0 b_0-c^2}{a b_0-c^2}},
\end{aligned}
\end{equation}
\noindent which relates to the single-detector counts by $\mu_{s} = \mu_T \Gamma_{s}$ and $\mu_{i} = \mu_T \Gamma_{i}$.\\
\indent The FHE, $\delta_{j}$, can then be approximated using the equations \cite{silberhorn}
\begin{equation}
\begin{gathered}
\delta_{s}=\frac{\Gamma_{\text {both }}}{\Gamma_i}=\sqrt{\frac{a_0 b-c^2}{a b-c^2}} \\
\delta_{i}=\frac{\Gamma_{\text {both }}}{\Gamma_s}=\sqrt{\frac{a b_0-c^2}{a b-c^2}} .
\end{gathered}
\end{equation}
\indent In the paper, we also investigate the spectral purity as a function of filtering to relate to the ability to perform HOM interference with independent photons. The equation for the single-mode purity for a heralded single photon, $\mathcal{P}$, is derived in refs.~\cite{JSI_model, silberhorn}. Using the above Gaussian approximation, this results in the equation \cite{silberhorn}
\begin{equation}
\mathcal{P} \approx \sqrt{\frac{a b-c^2}{a b}}.
\end{equation}

\end{document}